# Half-Quadratic Criterion based Adaptive Graph Signal Processing Algorithm

Chong Zhang[1],Haiquan Zhao[1*], Chengjin Li[1]

**Abstract :** In recent years, progress in adaptive graph signal processing algorithms has provided effective solutions for processing signals defined on graph structures. As a classical strategy in information theory, the Generalized Maximum Correntropy Criterion (GMCC) exhibits good resistance to non-Gaussian noises. When non-Gaussian noise interferes with the graph signal, the graph signal processing algorithm based on GMCC (GSP GMCC) algorithm shows better performance. However, the GSP GMCC algorithm itself has three parameters that need to be manually tuned, and the process of manually tuning the parameters is complex and tedious. Meanwhile, the non-concave and non-convex nature of the GMCC function itself limits its own convergence rate and adaptive estimation accuracy. To solve the above problems, based on the strongly convex function half-quadratic criterion (HQC), the GSP HQC algorithm is proposed in this paper. The performance analysis of the GSP HQC algorithm is implemented in this paper. Simulation experiments demonstrate that the GSP HQC algorithm achieves superior performance in terms of convergence rate and adaptive estimation accuracy while maintaining computational complexity comparable to existing algorithms.

**Keywords:** Adaptive estimation accuracy, convergence rate, Graph Signal Processing, Generalized Maximum Correntropy Criterion, half-quadratic criterion

## 1. Introduction:

In daily life, the common speech signal is defined in regular and neat one-dimensional domain, when it is disturbed by noise, the problem of degradation in the quality of speech communication will occur. In this regard, adaptive filtering algorithms provide an effective solution, and the related filtering algorithms are being widely used in scenarios such as active noise control (AEC) and acoustic echo cancellation (AEC) [3],[5],[7],[41],[26],[17],[52],[32],[53].

However, recent advances in science and technology have also led to increasingly complex data structures and higher dimensionality of data. In the aforementioned context, more and more data are distributed on irregular data structures such as social networks, sensor networks, big data networks and biological networks [33],[35],[31],[15],[23],[16]. To carry out denoising of these irregular data structures, adaptive filtering algorithms are inadequate, and the primary challenge is how to establish a connection between the classic adaptive filtering algorithms and the irregular data structures. To solve the above contradiction, under the influence of a technique known as graph theory, scholars involved in the signal processing community have successfully realized the extension of traditional signal processing techniques from regular domains to generalized graphs, thus proposing the graph signal processing (GSP). After years of efforts by the related scholars, GSP technology has been gradually applied to many fields, such as mechanical fault diagnosis, computer vision, multimedia processing and graph neural network [42],[43],[19],[44],[39],[27],[8],[9].Herein, graph signal adaptive processing algorithms can be employed as preprocessing or denoising layers for Graph Neural Networks (GNNs) operating in noisy environments [1],[2],[4],[6],[26],[20],[50], and they can compensate for the weaknesses of Graph Gaussian Processes (GPs) in terms of scalability with efficient and deterministic robust denoising [21].

[1] Chong Zhang, Haiquan Zhao, Chengjin Li are with the Key Laboratory of Magnetic Suspension Technology and Maglev Vehicle, Ministry of Education, and the School of Electrical Engineering, Southwest Jiaotong University, Chengdu610031, China.(email:chongzhang12@ 126.com; hqzhao_swjtu@126.com; chengjinliswjtu@126.com)

[*] Corresponding author: Haiquan Zhao.

In the above field, the study of adaptive graph signal processing algorithms provides possible solutions for graph signal denoising. Inspired by the classical least mean square (LMS) and normalized least mean square (NLMS) strategies, the LMS of the graph signal processing (GSP LMS) and the NLMS of the graph signal processing (GSP NLMS) algorithm have been successively proposed [14],[36]. When disturbed by non-Gaussian noise, there will be a serious deterioration or even divergence in the performance of the GSP LMS algorithm. As a classical strategy in the field of information theory, the Generalized Maximum Correntropy Criterion (GMCC) has a better performance against non-Gaussian noise [51],[54],[55],[12],[57], and the proposed GSP GMCC algorithm shows better robustness when non-Gaussian noise interferes with the graph signal [56]. Nevertheless, the manual debugging process for the three parameters of the GSP GMCC algorithm is complex and tedious. Simultaneously, the non-convex and non-concave characteristics of the GMCC algorithm itself constrain its performance in terms of convergence rate and adaptive estimation accuracy. For scenarios where the noise follows an alpha-stable distribution, researchers have proposed the Least Mean p-th for GSP (GSP LMP) algorithm and the Distributed Fractional-order Least Mean Squares (DFLMS) algorithm. However, GSP LMP requires computing the p-th power, and the optimal choice of p depends on the prior noise distribution [28]. While DFLMS achieves robust estimation of heavy-tailed noise over multi-agent networks, it relies on assumptions about noise parameters and network connectivity, requires additional parameter tuning (increasing computational overhead), and does not explicitly utilize GSP priors such as the graph spectrum or bandwidth constraints [38].To address the limitations of the above algorithms in terms of complexity, the Sign for GSP (GSP Sign) algorithm has been proposed. While the GSP Sign algorithm exhibits excellent performance and has low computational complexity in the presence of impulse noise [49], it suffers from lower steady-state accuracy under conditions of low impulse noise probability and performs poorly when confronted with other types of non-Gaussian noise.

To address the issues of overly complex parameter tuning in existing algorithms, limitations in performance imposed by the intrinsic properties of the functions themselves, and performance degradation in certain non-Gaussian noise environments, based on the half-quadratic criterion (HQC) function with strong convexity property [22],[29],[10],[48], the HQC cost function applicable to graph signals is constructed, and combined with stochastic gradient descent method, the GSP HQC algorithm is proposed. Furthermore, the related theoretical analyze of the GSP HQC algorithm is performed, which will determine the range of step size limits and the steady state case of the GSP HQC algorithm. Finally, the superior performance of the GSP HQC algorithm in terms of convergence rate and adaptive estimation accuracy is verified by performing computer simulation experiments.

The contribution of our work to the field of graph signal processing algorithms can be summarized as follows:

1) Based on the HQC function with strongly convex nature, the GSP HQC algorithm is proposed in this paper, which has one less parameter to be debugged compared to the GSP GMCC algorithm.

2) The corresponding theoretical analysis of the GSP HQC algorithm is carried out in this paper, which will determine the range of step size limits and the steady state case of the GSP HQC algorithm.

3) By performing computer simulation experiments, the convergence rate advantage and adaptive estimation accuracy advantage of the GSP HQC algorithm are effectively demonstrated.

The paper is structured as follows: The basics of HQC and graph signal processing are described in detail in Section II. The derivation of the GSP HQC algorithm will be completed in Section III. The related theoretical analysis of the GSP HQC algorithm will be given in Section IV. The computer simulations will be performed in Section V. Finally, the summary of the entire paper will be presented in Section VI.

## 2. GSP HQC derivation

### 2.1 The basic knowledge of Graph Signal Processing

A graph $\mathbf{G}(\mathbf{V},\varepsilon)$ is composed of a set of nodes $\mathbf{V}=\{v_1,v_2,v_3,...,v_N\}$, which are connected to each other by edges $\varepsilon$. $\{v_m,v_n\}$ is used to denote node $v_m\in\mathbf{V}$ that are connected to node $v_n\in\mathbf{V}$ to form an edge. In graph theory, in order to characterize the strength of an edge formed by the connection of two nodes, each edge of the graph $\mathbf{G}(\mathbf{V},\varepsilon)$ is associated with a weight variable. The weight matrix consisting of the weights of N nodes is called the adjacency matrix $\mathbf{A}\in\mathbf{R}^{N\times N}$ and is used to characterize the structure of the graph. In graph theory, graphs are classified into directed and undirected graphs based on the existence of bidirectional connections between nodes, and the adjacency matrix of an undirected graph is a symmetric matrix. The elements of the mth row and nth column of the matrix $a_{mn}$ represent the strength of the connections between node $v_m\in\mathbf{V}$ and node $v_n\in\mathbf{V}$. Note that if there is no edge connection between two nodes, the corresponding weight factor of the edge is zero.

For a graph $\mathbf{G}$ with N nodes, a signal defined on is called a graph signal, which is denoted by $\mathbf{x} \in \mathbf{R}^{N \times 1}$. The nth component $x_n$ of the graph signal $\mathbf{x}$ represents the signal value of the nth node on the graph $\mathbf{G}$. Driven by the classical signal processing framework, the concept of Graph Fourier Transform (GFT) [40],[13] was introduced. With GFT, the spectral decomposition of the adjacency matrix $\mathbf{A}$ has been implemented. Considering the case of undirected graphs, the spectral decomposition of the adjacency matrix is shown in (1)

$$\mathbf{A} = \mathbf{U\Pi U}^T \tag{1}$$

where $\mathbf{\Pi} = \mathrm{diag}\{\lambda_1, \lambda_2, ..., \lambda_N\}$ is a diagonal matrix consisting of the eigenvalues of the adjacency matrix $\mathbf{A}$, and $\mathbf{U} \in \mathbf{R}^{N \times N}$ is a matrix consisting of the set of eigenvectors $\{\mathbf{u}_i\}_{i \in N}$ of the matrix $\mathbf{A}$. The expression of $\mathbf{U}$, and the relationship between $\{\mathbf{u}_i\}_{i \in N}$, $\lambda_i$ and $\mathbf{A}$ is shown in (2)

$$\begin{cases} \mathbf{U} = [\mathbf{u}_1, \mathbf{u}_2, \mathbf{u}_3, ..., \mathbf{u}_N] \\ \mathbf{A}\mathbf{u}_i = \lambda_i \mathbf{u}_i \qquad i \in N \end{cases} \tag{2}$$

In this paper, the case is only considered where $\mathbf{A}$ is decomposable.

The defining of GFT is presented in (3)

$$\mathbf{s} = \mathbf{U}^T \mathbf{x} \tag{3}$$

The defining equation of inverse graph Fourier transform (IGFT) is presented in (4)

$$\mathbf{x} = \mathbf{U}\mathbf{s} \tag{4}$$

A graph signal $\mathbf{x}$ is said to be frequency-domain sparse or bandwidth-limited when there are zeros in the components of the frequency-domain signal s. For a subset F of the nodal domain $\mathbf{V}$, if the graph signal is spectrally sparse, the frequency-domain signal satisfies (5)

$$\mathbf{F} = \{ \exists p \in \mathbf{V} \,|\, \mathrm{s}_p \neq 0 \} \tag{5}$$

Under the premise that the graph signal is F-spectrally sparse, the graph signal can be expressed in the form of (6)

$$\mathbf{x} = \mathbf{U}_F \mathbf{s}_F \tag{6}$$

where $\mathbf{U}_F \in \mathbf{R}^{N \times F}$ is the matrix consisting of $\{\mathbf{u}_m\}_{m \in \mathbf{F}}$ and $\mathbf{s}_F \in \mathbf{R}^{F \times 1}$ is the frequency domain expression of the graph signal obtained by crossing out the components with a value of 0 from the subset F.

Based on the relevant practical needs of saving resources, the concept of sampling matrix is proposed, multiplying the original graph signal $\mathbf{x}$ and the sampling matrix $\mathbf{D}_{\mathrm{S}} \in \mathbf{R}^{N\times N}$ [13],[24], the actual graph sampling signal is obtained. $\mathbf{D}_{\mathrm{S}}$ is a diagonal matrix consisting only of 0 or 1, and its corresponding element is 1 when the data on the node can be acquired, otherwise it is 0. The graph sampling signal $\mathbf{x}_{\mathrm{s}} \in \mathbf{R}^{N\times 1}$ is shown in (7)

$$\mathbf{x}_{\mathrm{s}} = \mathbf{D}_{\mathrm{S}}\mathbf{x} \tag{7}$$

## 2.2 The HQC function

By defining $e(i) = x(i) - y(i)$ and introducing the design parameter $\tau$, the HQC cost function is shown in (8)

$$J_{\mathrm{HQC}}\left(e(i)\right) = \frac{1}{\tau}\left\{\mathrm{E}\left[\sqrt{1+\tau e^{2}(i)}\right] - 1\right\} \tag{8}$$

The variation of HQC and its derivative function with error for different values of the parameters are shown in Fig. 1 and Fig. 2, respectively.

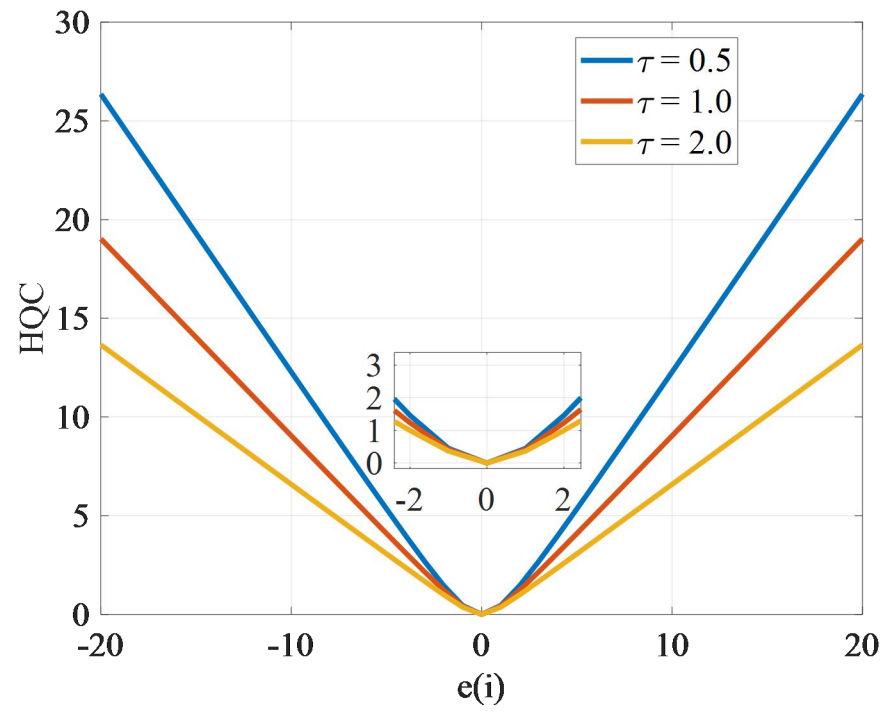

Fig. 1. Variation of HQC with error

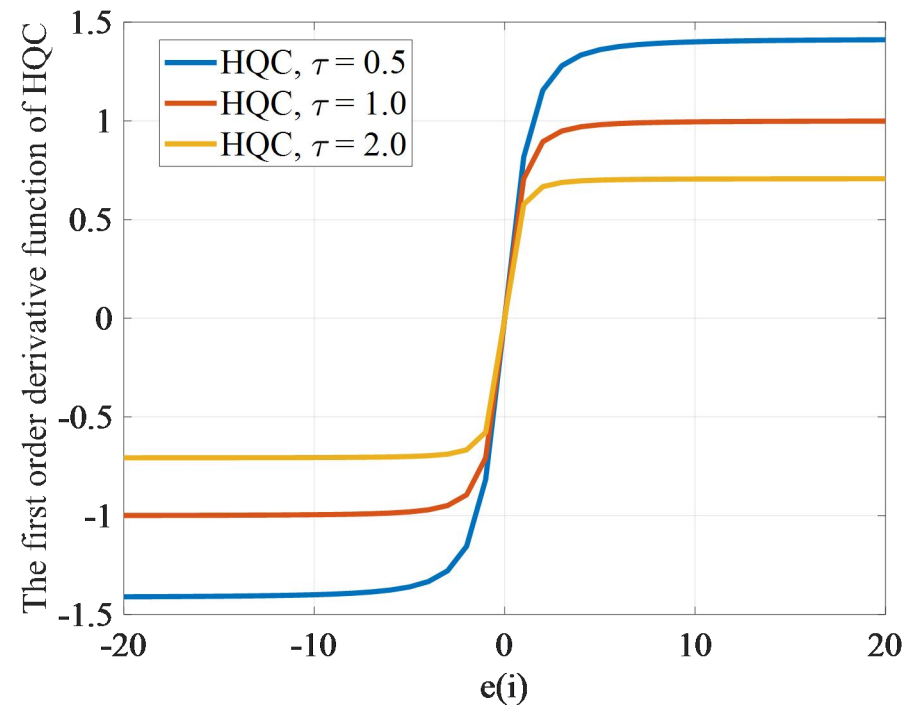

Fig. 2. Variation of the first-order derivative function of HQC with error

From Fig. 1, it can be found that the HQC has a quadratic (MSE) shape when the error value is small and a linear (Mean Absolute Error, MAE) shape when the error value is large.

From Fig. 2, it can be noticed that unlike the LOG [45] and MVC [18],[11] cost functions which are non-convex (non-positive definite) for all error values, the HQC itself is a convex function whose first-order derivative is differentiable and bounded, which is also verified in Fig. 2. Since any local minimum of a convex function is also a global minimum, many optimization algorithms are guaranteed to converge to a global minimum on convex functions, and therefore, convex functions have better stability than non-convex and concave functions. The above properties ensure better robust performance when using HQC based adaptive signal processing algorithms.

For convenience, we investigate the respective Hessian matrix coefficients of the LOG function,

which is similar to the HQC function and also has only one argument. Both the LOG function and the GMCC function are nonconcave and nonconvex functions. In addition, existing work has shown that the results of the derivation of the HQC function compared to the results of the derivation of the LOG function are only differentiated by the extra root sign. The Hessian matrix coefficients [46] of the LOG and HQC functions are shown in (9) and (10), respectively.

$$\gamma_i^{\mathrm{LOG}} = \frac{2\tau\left(1-\tau e^2\left(i\right)\right)}{\left(1+\tau\left|e\left(i\right)\right|^2\right)^2} \tag{9}$$

$$\gamma_i^{\mathrm{HQC}} = \frac{1}{\sqrt{\left(1+\tau e^2\left(i\right)\right)^3}} \tag{10}$$

The variation of the Hessian matrix coefficients of LOG and HQC with error when $\tau$ takes different values is shown in Fig. 3.

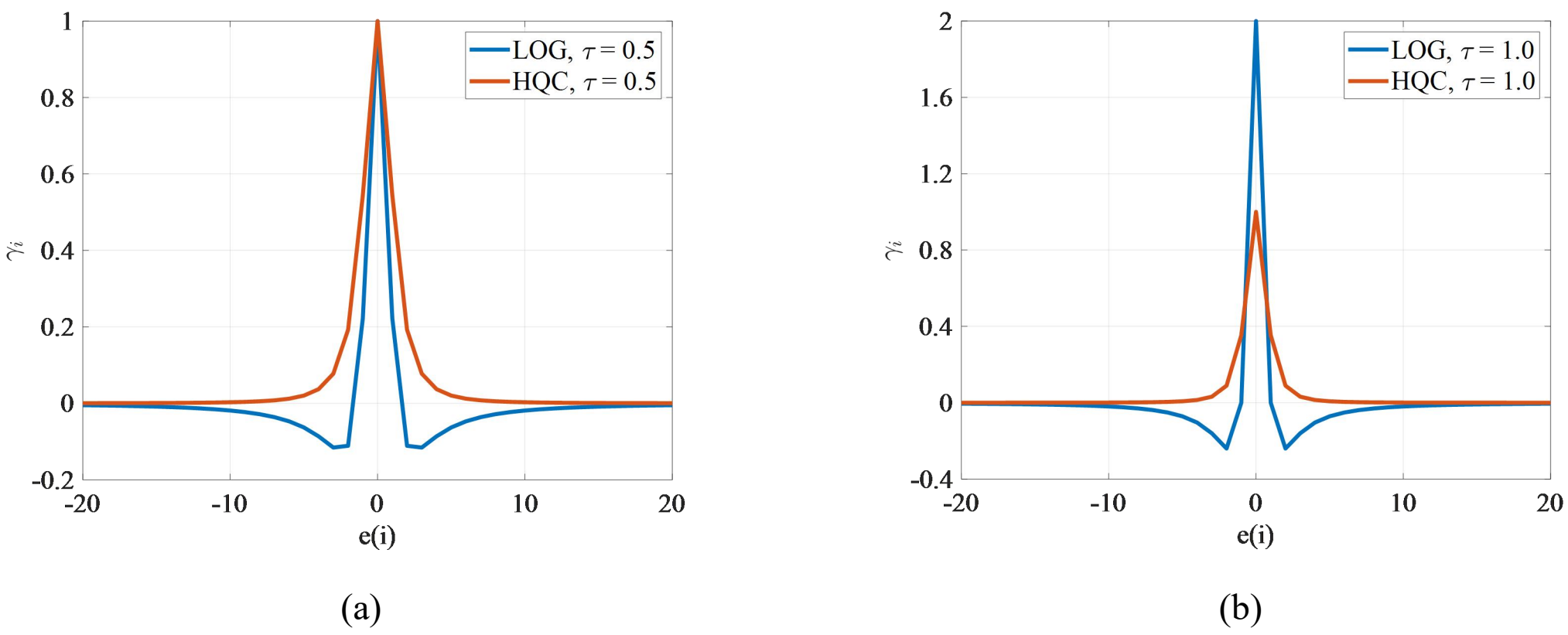

Fig. 3. Variation of Hessian matrix coefficients with error for LOG and HQC: (a) $\tau = 0.5$ (b) $\tau = 1.0$

From (9), (10) and Fig. 3, it can be found that when $\left|e\left(i\right)\right| < \sqrt{1/\tau}$ , the LOG function is a convex function, when $e\left(i\right)$ is in the other range, the LOG function is a nonconvex function. In contrast, the fact that the Hessian matrix coefficients of the HQC function are positive in all cases ensures the strongly convex nature of the HQC function.

In the following, this paper investigates the performance surfaces of the MSE and HQC functions in a specific scenario where the input signal of the adaptive filtering algorithm is disturbed by noise, the performance surface of HQC is obtained by varying the value of $\tau$ in each of the scenario, and they are compared with the corresponding MSE performance surface. The scenarios are shown as follows:

Scenario: The input signal has a mean of 0 and a variance of 5, the noise signal has a mean of 0 and a variance of 0.8, and the signal length is 500. The true weight is $\mathbf{w}_{\mathrm{o}} = \left[6.8, 3.2\right]^T$ , and the performance surface results are shown in Fig. 4.

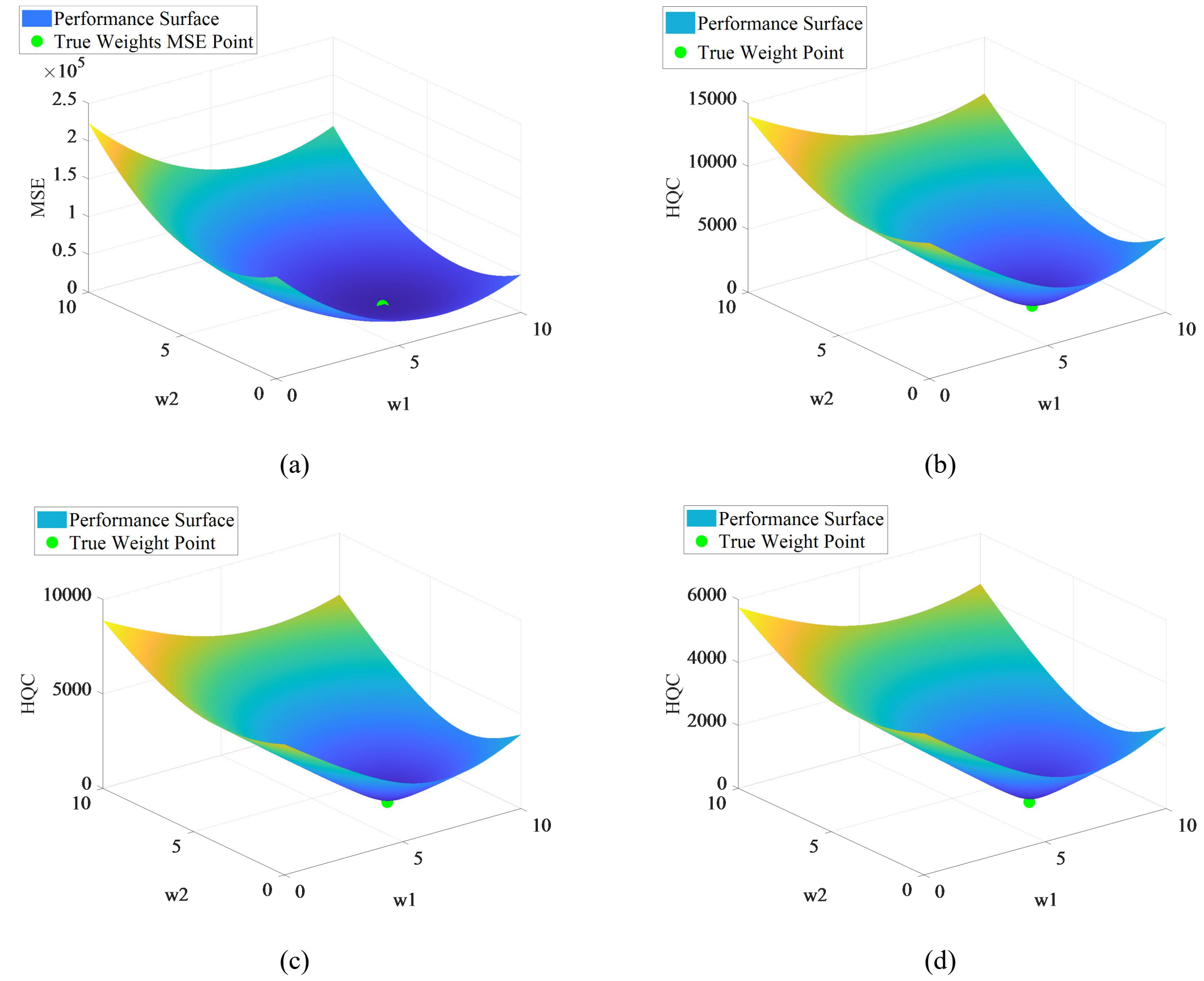

Fig. 4. Performance Surface Results: (a) MSE (b) HQC, $\tau = 0.3$ (c) HQC, $\tau = 0.8$ (d) HQC, $\tau = 2.0$

From the results in Fig. 4, it can be found that when larger outliers occur, the gradient of the HQC will gradually decrease to a very significant degree, which effectively mitigates the effect of larger outliers on the drastic perturbation of the adaptive filter coefficients. Therefore, in the same non-Gaussian noise environment, the HQC-based adaptive signal processing algorithm has higher adaptive estimation accuracy and better robustness.

## 2.3 The proposal of the GSP HQC algorithm

Considering that only impulsive noise among the current non-Gaussian noise has a clear and explicit nature, this paper proposes that the noise background of the GSP HQC algorithm and its related performance analysis are all impulsive noise.

When the impulse noise interferes with the $\mathbf{x}_o$, the noisy graph signal at moment i is shown in (11)

$$\begin{aligned} \mathbf{x}_w(i) &= \mathbf{x}_o + \mathbf{w}(i) \\ &= \mathbf{x}_o + \boldsymbol{\eta}(i) + \mathbf{B}\boldsymbol{\gamma}(i) \end{aligned} \tag{11}$$

where $\boldsymbol{\eta}(i)$ is Gaussian noise signal with mean 0 and variance $\delta_\eta^2$, $\boldsymbol{\gamma}(i)$ is Gaussian noise signal with mean 0 and variance $\delta_\gamma^2$ , and $\delta_\gamma^2 >> \delta_\eta^2$ . B is a diagonal matrix generated by the Bernoulli process whose elements are only 0 or 1 and satisfies (12)

$$\begin{cases} P\left(b_{ii}=1\right)=Pr \\ P\left(b_{ii}=0\right)=1-Pr \end{cases} \tag{12}$$

where Pr is the probability of the occurrence of impulse noise, and $b_{ii}$ is the element in the ith row and ith column of $\mathbf{B}$ .

The established cost function of the GSP HQC algorithm is shown in (13)

$$J_{\text{GSP HQC}}\left(\hat{\mathbf{s}}_F\left(i\right)\right)=\frac{1}{\tau}\left[\sqrt{1+\tau e_1^2\left(i\right)}-1\right]+\frac{1}{\tau}\left[\sqrt{1+\tau e_2^2\left(i\right)}-1\right]+\ldots+\frac{1}{\tau}\left[\sqrt{1+\tau e_N^2\left(i\right)}-1\right] \tag{13}$$

where $\tau$ is the design parameter of $J_{\text{GSP HQC}}\left(\hat{\mathbf{s}}_F\left(i\right)\right)$ and $e_k^2\left(i\right)$ is the squared value of the kth component of the error vector at the moment i.

Taking the gradient of $J_{\text{GSP HQC}}\left(\hat{\mathbf{s}}_F\left(i\right)\right)$, it can be obtained that

$$\begin{aligned}\nabla J_{\text{GSP HQC}}\left(\hat{\mathbf{s}}_F\left(i\right)\right)&=\frac{\partial J_{\text{GSP HQC}}}{\partial \hat{\mathbf{s}}_F\left(i\right)}\\ &=-\mathbf{U}_F^T\times\begin{bmatrix}\frac{1}{\sqrt{1+\tau e_1^2\left(i\right)}} & 0 & 0 & \cdots & 0\\ 0 & \frac{1}{\sqrt{1+\tau e_2^2\left(i\right)}} & 0 & 0 & \cdots\\ 0 & 0 & \frac{1}{\sqrt{1+\tau e_3^2\left(i\right)}} & & \cdots\\ \vdots & \vdots & \vdots & & \cdots\\ 0 & 0 & \cdots & & \frac{1}{\sqrt{1+\tau e_N^2\left(i\right)}}\end{bmatrix}\times\begin{bmatrix}e_1\left(i\right)\\ e_2\left(i\right)\\ e_3\left(i\right)\\ \vdots\\ e_N\left(i\right)\end{bmatrix}\end{aligned} \tag{14}$$

Since $\hat{\mathbf{x}}_{\mathbf{o}}\left(i\right)=\mathbf{U}_F\hat{\mathbf{s}}_F\left(i\right)$, the GSP HQC algorithm is shown in (15)

$$\begin{aligned}\hat{\mathbf{x}}_{\mathbf{o}}\left(i+1\right)&=\hat{\mathbf{x}}_{\mathbf{o}}\left(i\right)-\mu\mathbf{U}_F\left[-\mathbf{U}_F^T\mathbf{G}\left(\mathbf{e}\left(i\right)\right)\mathbf{e}\left(i\right)\right]\\ &=\hat{\mathbf{x}}_{\mathbf{o}}\left(i\right)+\mu\mathbf{U}_F\mathbf{U}_F^T\mathbf{G}\left(\mathbf{e}\left(i\right)\right)\mathbf{e}\left(i\right)\end{aligned} \tag{15}$$

The expression of $\mathbf{G}\left(\mathbf{e}\left(i\right)\right)$ can be found in (16)

$$\mathbf{G}\left(\mathbf{e}(i)\right)=\begin{bmatrix} \frac{1}{\sqrt{1+\tau e_1^2(i)}} & 0 & 0 & \cdots & 0 \\ 0 & \frac{1}{\sqrt{1+\tau e_2^2(i)}} & 0 & 0 & \cdots \\ 0 & 0 & \frac{1}{\sqrt{1+\tau e_3^2(i)}} & & \cdots \\ \vdots & \vdots & \vdots & & \cdots \\ 0 & 0 & \cdots & & \frac{1}{\sqrt{1+\tau e_N^2(i)}} \end{bmatrix} \tag{16}$$

From (16), it can be found that when non-Gaussian noise, such as impulse noise, occurs, $\mathbf{G}\left(\mathbf{e}(i)\right)$ can be made to play a better role in suppressing larger outliers by adjusting the value of $\tau$.

Remark: It should be noted that in the second section of this paper, the LOG function is used to compare with the HQC function to highlight the strongly convex nature of the HQC function. The error handling matrix of the GSP LOG algorithm derived based on the LOG function only lacks the difference of square root compared to $\mathbf{G}\left(\mathbf{e}(i)\right)$.

## 3. Theoretical performance analysis

The theoretical performance analysis of the proposed GSP HQC algorithm will be performed in this section, including mean stability analysis, mean square stability analysis, steady state error analysis and computational complexity analysis.

$\tilde{\mathbf{s}}_F(i)$ is defined in (17)

$$\tilde{\mathbf{s}}_F(i)=\mathbf{s}_F-\hat{\mathbf{s}}_F(i) \tag{17}$$

The following reasonable assumptions have been adopted in this section for the convenience and correct conduct of the theoretical performance analysis:

**Assumption1:** The graph signal x is bandwidth limited.

**Assumption2:** The sampling matrix corresponds to a sampling set that is constant, i.e. $\mathbf{D_S}(i)=\mathbf{D_S}$.

**Assumption3:** Under the small-step-size hypothesis, the slowly varying $\mathbf{G}\left(\mathbf{e}(i)\right)$ under steady-state conditions is replaced by its steady-state counterpart, i.e., the error handling matrix $\mathbf{G}\left(\mathbf{e}(i)\right)$ is a constant matrix [34].

**Assumption4:** The noise signals $\boldsymbol{\eta}(i)$ and $\boldsymbol{\gamma}(i)$ are Gaussian noises that satisfy the same distribution with zero mean and the components are independent and uncorrelated with each other.

**Assumption5:** $\mathbf{U}_F$, $\mathbf{D}_\mathrm{S}$, $\mathbf{G}(\mathbf{e}(i))$, $\boldsymbol{\eta}(i)$, $\boldsymbol{\gamma}(i)$ and $\tilde{\mathbf{s}}_F(i)$ are independent and unrelated to each other.

**Assumption6:**Employing the independence assumption, the statistical coupling terms are neglected, meaning the weight error vector and the input signal are statistically independent [34].

## 3.1 Mean stability analysis

Based on the frequency domain update formula in (14) and the frequency domain error vector at the moment i shown in (17), the frequency domain error vector at the moment (i+1) can be obtained, as shown in (18)

$$\begin{aligned}
\tilde{\mathbf{s}}_F(i+1) &= \mathbf{s}_F - \hat{\mathbf{s}}_F(i+1) \\
&= \mathbf{s}_F - \hat{\mathbf{s}}_F(i) - \mu \mathbf{U}_F^T \mathbf{G}(\mathbf{e}(i))\mathbf{e}(i) \\
&= \tilde{\mathbf{s}}_F(i) - \mu \mathbf{U}_F^T \mathbf{G}(\mathbf{e}(i))\mathbf{D}_\mathrm{S}\left[\mathbf{x}_\mathbf{o} + \mathbf{w}(i) - \hat{\mathbf{x}}_\mathbf{o}(i)\right] \\
&= \tilde{\mathbf{s}}_F(i) - \mu \mathbf{U}_F^T \mathbf{G}(\mathbf{e}(i))\mathbf{D}_\mathrm{S}\left[\mathbf{U}_F\mathbf{s}_F + \mathbf{w}(i) - \mathbf{U}_F\hat{\mathbf{s}}_F(i)\right] \\
&= \tilde{\mathbf{s}}_F(i) - \mu \mathbf{U}_F^T \mathbf{G}(\mathbf{e}(i))\left[\mathbf{D}_\mathrm{S}\mathbf{U}_F\tilde{\mathbf{s}}_F(i) + \mathbf{D}_\mathrm{S}\mathbf{w}(i)\right] \\
&= \left[\mathbf{I} - \mu \mathbf{U}_F^T \mathbf{G}(\mathbf{e}(i))\mathbf{D}_\mathrm{S}\mathbf{U}_F\right]\tilde{\mathbf{s}}_F(i) - \mu \mathbf{U}_F^T \mathbf{G}(\mathbf{e}(i))\mathbf{D}_\mathrm{S}\mathbf{w}(i)
\end{aligned} \tag{18}$$

Taking the mathematical expectation on both sides of (18), it can be obtained that

$$\mathrm{E}\left[\tilde{\mathbf{s}}_F(i+1)\right] = \mathrm{E}\left[\mathbf{I} - \mu \mathbf{U}_F^T \mathbf{G}(\mathbf{e}(i))\mathbf{D}_\mathrm{S}\mathbf{U}_F\right]\mathrm{E}\left[\tilde{\mathbf{s}}_F(i)\right] - \mu \mathrm{E}\left[\mathbf{U}_F^T \mathbf{G}(\mathbf{e}(i))\mathbf{D}_\mathrm{S}\mathbf{w}(i)\right] \tag{19}$$

Based on Assumption 4 and Assumption 5, one can conclude that

$$\begin{aligned}
\mathrm{E}\left[\tilde{\mathbf{s}}_F(i+1)\right] &= \mathrm{E}\left[\mathbf{I} - \mu \mathbf{U}_F^T \mathbf{G}(\mathbf{e}(i))\mathbf{D}_\mathrm{S}\mathbf{U}_F\right]\mathrm{E}\left[\tilde{\mathbf{s}}_F(i)\right] - \mu \mathrm{E}\left[\mathbf{U}_F^T \mathbf{G}(\mathbf{e}(i))\mathbf{D}_\mathrm{S}\right]\mathrm{E}\left[\mathbf{w}(i)\right] \\
&= \left[\mathbf{I} - \mu \mathbf{U}_F^T \mathbf{G}(\mathbf{e}(i))\mathbf{D}_\mathrm{S}\mathbf{U}_F\right]\mathrm{E}\left[\tilde{\mathbf{s}}_F(i)\right] \\
&\quad - \mu \mathbf{U}_F^T \mathbf{G}(\mathbf{e}(i))\mathbf{D}_\mathrm{S}\left(\mathrm{E}\left[\mathbf{q}(i)\right] + \mathrm{E}\left[\mathbf{B}\right]\mathrm{E}\left[\mathbf{p}(i)\right]\right) \\
&= \left[\mathbf{I} - \mu \mathbf{U}_F^T \mathbf{G}(\mathbf{e}(i))\mathbf{D}_\mathrm{S}\mathbf{U}_F\right]\mathrm{E}\left[\tilde{\mathbf{s}}_F(i)\right] - \mu \mathbf{U}_F^T \mathbf{G}(\mathbf{e}(i))\mathbf{D}_\mathrm{S}\left(\mathbf{0} \in \mathbf{R}^{N\times 1}\right) \\
&= \left[\mathbf{I} - \mu \mathbf{U}_F^T \mathbf{G}(\mathbf{e}(i))\mathbf{D}_\mathrm{S}\mathbf{U}_F\right]\mathrm{E}\left[\tilde{\mathbf{s}}_F(i)\right]
\end{aligned} \tag{20}$$

Since both $\mathbf{D}_\mathrm{S}$ and $\mathbf{G}(\mathbf{e}(i))$ are diagonal matrices, $\mathbf{D}_\mathrm{S}$ and $\mathbf{G}(\mathbf{e}(i))$ are exchangeable, i.e. $\mathbf{D}_\mathrm{S}\mathbf{G}(\mathbf{e}(i)) = \mathbf{G}(\mathbf{e}(i))\mathbf{D}_\mathrm{S}$, it can be found that

$$\left[\mathbf{U}_F^T \mathbf{G}(\mathbf{e}(i))\mathbf{D}_\mathrm{S}\mathbf{U}_F\right]^T = \mathbf{U}_F^T \mathbf{D}_\mathrm{S}\mathbf{G}(\mathbf{e}(i))\mathbf{U}_F = \mathbf{U}_F^T \mathbf{G}(\mathbf{e}(i))\mathbf{D}_\mathrm{S}\mathbf{U}_F \tag{21}$$

The establishment of (21) shows that $\mathbf{U}_F^T\mathbf{G}\left(\mathbf{e}(i)\right)\mathbf{D}_\mathbf{S}\mathbf{U}_F$ is a symmetric matrix, then there exists a spectral decomposition form of $\mathbf{U}_F^T\mathbf{G}\left(\mathbf{e}(i)\right)\mathbf{D}_\mathbf{S}\mathbf{U}_F$, which is shown in (22)

$$\mathbf{U}_F^T\mathbf{G}\left(\mathbf{e}(i)\right)\mathbf{D}_\mathbf{S}\mathbf{U}_F = \mathbf{K}\boldsymbol{\Pi}\mathbf{K}^T \tag{22}$$

where $\boldsymbol{\Pi}$ is a diagonal matrix consisting of the eigenvalues of $\mathbf{U}_F^T\mathbf{G}\left(\mathbf{e}(i)\right)\mathbf{D}_\mathbf{S}\mathbf{U}_F$. $\mathbf{K}\in\mathbf{R}^{N\times N}$ is the corresponding orthogonal matrix, which satisfies $\mathbf{K}\mathbf{K}^T = \mathbf{K}^T\mathbf{K} = \mathbf{I}$.

Based on (20) and (22), it can be obtained that

$$\begin{aligned}\mathrm{E}\left[\tilde{\mathbf{s}}_F(i+1)\right] &= \left[\mathbf{I}-\mu\mathbf{U}_F^T\mathbf{G}\left(\mathbf{e}(i)\right)\mathbf{D}_\mathbf{S}\mathbf{U}_F\right]\mathrm{E}\left[\tilde{\mathbf{s}}_F(i)\right] \\ &= \left[\mathbf{K}\mathbf{I}\mathbf{K}^T-\mu\mathbf{K}\Pi\mathbf{K}^T\right]\mathrm{E}\left[\tilde{\mathbf{s}}_F(i)\right] \\ &= \mathbf{K}\left[\mathbf{I}-\mu\Pi\right]\mathbf{K}^T\mathrm{E}\left[\tilde{\mathbf{s}}_F(i)\right]\end{aligned} \tag{23}$$

Multiply both sides of equation (23) by $\mathbf{K}^T$, and make $\boldsymbol{\chi}(i)=\mathbf{K}^T\mathrm{E}\left[\tilde{\mathbf{s}}_F(i)\right]\in\mathbf{R}^{N\times 1}$, it can be found that

$$\begin{aligned}\boldsymbol{\chi}(i+1) &= \mathbf{K}^T\mathbf{K}\left[\mathbf{I}-\mu\boldsymbol{\Pi}\right]\mathbf{K}^T\mathrm{E}\left[\tilde{\mathbf{s}}_F(i)\right] \\ &= \left[\mathbf{I}-\mu\boldsymbol{\Pi}\right]\boldsymbol{\chi}(i)\end{aligned} \tag{24}$$

According to (24), to ensure that the GSP HQC algorithm remains convergence-stable in the mean sense, the magnitude of each component in $\boldsymbol{\chi}(i)$ has to be less than the magnitude of the corresponding component in $\boldsymbol{\chi}(i+1)$. Therefore, (25) must be satisfied.

$$\left|1-\lambda_{\max}\left(\mathbf{U}_F^T\mathbf{G}\left(\mathbf{e}(i)\right)\mathbf{D}_\mathbf{S}\mathbf{U}_F\right)\right|<1 \tag{25}$$

The range of step-size values that ensure mean stability of the GSP - HQC algorithm can be obtained from (25); specifically, for stability we require $\left|1-\mu\lambda\right|<1$ for every eigenvalue $\lambda$ of the matrix $\mathbf{U}_F^T\mathbf{G}\left(\mathbf{e}(i)\right)\mathbf{D}_\mathbf{S}\mathbf{U}_F$, as stated in (26)

$$0<\mu<\frac{2}{\lambda_{\max}\left(\mathbf{U}_F^T\mathbf{G}\left(\mathbf{e}(i)\right)\mathbf{D}_\mathbf{S}\mathbf{U}_F\right)} \tag{26}$$

## 3.2 Mean square stability analysis

Multiplying equation (18) by its transposed form and taking the mathematical expectation operator on both sides of the equation resulting from the multiplication of the first two, the relationship between $\mathrm{E}\left[\tilde{\mathbf{s}}_F(i+1)\tilde{\mathbf{s}}_F^T(i+1)\right]$ and $\mathrm{E}\left[\tilde{\mathbf{s}}_F(i)\tilde{\mathbf{s}}_F^T(i)\right]$ can be obtained, which can be found in (27)

$$
\begin{aligned}
&\mathrm{E}\left[\tilde{\mathbf{s}}_F(i+1)\tilde{\mathbf{s}}_F^T(i+1)\right] \\
&= \mathrm{E}\left\{\left[\mathbf{I}-\mu\mathbf{U}_F^T\mathbf{G}(\mathbf{e}(i))\mathbf{D}_S\mathbf{U}_F\right]\tilde{\mathbf{s}}_F(i) - \mathbf{U}_F^T\mathbf{G}(\mathbf{e}(i))\mathbf{D}_S\mathbf{w}(i)\right\} \\
&\quad \times \mathrm{E}\left\{\left[\mathbf{I}-\mu\mathbf{U}_F^T\mathbf{D}_S\mathbf{G}(\mathbf{e}(i))\mathbf{U}_F\right]\tilde{\mathbf{s}}_F^T(i) - \mathbf{w}^T(n)\mathbf{D}_S\mathbf{G}(\mathbf{e}(i))\mathbf{U}_F\right\} \\
&= \mathrm{E}\left\{\left[\mathbf{I}-\mu\mathbf{U}_F^T\mathbf{G}(\mathbf{e}(i))\mathbf{D}_S\mathbf{U}_F\right]\tilde{\mathbf{s}}_F(i)\tilde{\mathbf{s}}_F^T(i)\left[\mathbf{I}-\mu\mathbf{U}_F^T\mathbf{D}_S\mathbf{G}(\mathbf{e}(i))\mathbf{U}_F\right]\right\} \\
&\quad -\mu\mathrm{E}\left\{\mathbf{U}_F^T\mathbf{G}(\mathbf{e}(i))\mathbf{D}_S\mathbf{w}(i)\tilde{\mathbf{s}}_F^T(i)\left[\mathbf{I}-\mu\mathbf{U}_F^T\mathbf{D}_S\mathbf{G}(\mathbf{e}(i))\mathbf{U}_F\right]\right\} \\
&\quad -\mu\mathrm{E}\left\{\left[\mathbf{I}-\mu\mathbf{U}_F^T\mathbf{G}(\mathbf{e}(i))\mathbf{D}_S\mathbf{U}_F\right]\tilde{\mathbf{s}}_F(i)\mathbf{w}^T(i)\mathbf{D}_S\mathbf{G}(\mathbf{e}(i))\mathbf{U}_F\right\} \\
&\quad +\mu^2\mathrm{vec}\left\{\mathrm{E}\left[\mathbf{U}_F^T\mathbf{G}(\mathbf{e}(i))\mathbf{D}_S\mathbf{w}(i)\mathbf{w}^T(i)\mathbf{D}_S\mathbf{G}(\mathbf{e}(i))\mathbf{U}_F\right]\right\}
\end{aligned} \tag{27}
$$

Based on Assumption4 and Assumption5, (28) and (29) can be obtained respectively.

$$
\mu\mathrm{E}\left\{\mathbf{U}_F^T\mathbf{G}(\mathbf{e}(i))\mathbf{D}_S\mathbf{w}(i)\tilde{\mathbf{s}}_F^T(i)\left[\mathbf{I}-\mu\mathbf{U}_F^T\mathbf{D}_S\mathbf{G}(\mathbf{e}(i))\mathbf{U}_F\right]\right\} = \mathbf{0} \in \mathbf{R}^{F\times F} \tag{28}
$$

$$
\mu\mathrm{E}\left\{\left[\mathbf{I}-\mu\mathbf{U}_F^T\mathbf{G}(\mathbf{e}(i))\mathbf{D}_S\mathbf{U}_F\right]\tilde{\mathbf{s}}_F(i)\mathbf{w}^T(i)\mathbf{D}_S\mathbf{G}(\mathbf{e}(i))\mathbf{U}_F\right\} = \mathbf{0} \in \mathbf{R}^{F\times F} \tag{29}
$$

Based on (28) and (29), (27) can be simplified to (30)

$$
\begin{aligned}
&\mathrm{E}\left[\tilde{\mathbf{s}}_F(i+1)\tilde{\mathbf{s}}_F^T(i+1)\right] \\
&= \mathrm{E}\left\{\left[\mathbf{I}-\mu\mathbf{U}_F^T\mathbf{G}(\mathbf{e}(i))\mathbf{D}_S\mathbf{U}_F\right]\tilde{\mathbf{s}}_F(i)\tilde{\mathbf{s}}_F^T(i)\left[\mathbf{I}-\mu\mathbf{U}_F^T\mathbf{D}_S\mathbf{G}(\mathbf{e}(i))\mathbf{U}_F\right]\right\} \\
&\quad +\mu^2\mathrm{E}\left[\mathbf{U}_F^T\mathbf{G}(\mathbf{e}(i))\mathbf{D}_S\mathbf{w}(i)\mathbf{w}^T(i)\mathbf{D}_S\mathbf{G}(\mathbf{e}(i))\mathbf{U}_F\right]
\end{aligned} \tag{30}
$$

Processing (30) based on $\mathrm{vec}(\mathbf{ABC}) = (\mathbf{C}^T \otimes \mathbf{A})\,\mathrm{vec}(\mathbf{B})$ yields

$$
\begin{aligned}
&\mathrm{vec}\left\{\mathrm{E}\left[\tilde{\mathbf{s}}_F(i+1)\tilde{\mathbf{s}}_F^T(i+1)\right]\right\} \\
&= \left[\mathbf{I}-\mu\mathbf{U}_F^T\mathbf{G}(\mathbf{e}(i))\mathbf{D}_S\mathbf{U}_F\right]\otimes\left[\mathbf{I}-\mu\mathbf{U}_F^T\mathbf{G}(\mathbf{e}(i))\mathbf{D}_S\mathbf{U}_F\right]\mathrm{vec}\left\{\mathrm{E}\left[\tilde{\mathbf{s}}_F(i)\tilde{\mathbf{s}}_F^T(i)\right]\right\} \\
&\quad +\mu^2\mathrm{vec}\left\{\mathrm{E}\left[\mathbf{U}_F^T\mathbf{G}(\mathbf{e}(i))\mathbf{D}_S\mathbf{w}(i)\mathbf{w}^T(i)\mathbf{D}_S\mathbf{G}(\mathbf{e}(i))\mathbf{U}_F\right]\right\}
\end{aligned} \tag{31}
$$

Processing (31) based on $(\mathbf{A}+\mathbf{B})\otimes(\mathbf{C}+\mathbf{D}) = \mathbf{A}\otimes\mathbf{C}+\mathbf{B}\otimes\mathbf{C}+\mathbf{A}\otimes\mathbf{D}+\mathbf{B}\otimes\mathbf{D}$ yields

$$
\begin{aligned}
&\mathrm{vec}\left\{\mathrm{E}\left[\tilde{\mathbf{s}}_F(i+1)\tilde{\mathbf{s}}_F^T(i+1)\right]\right\} \\
&= \begin{bmatrix} \mathbf{I}\otimes\mathbf{I}-\mathbf{I}\otimes\mathbf{U}_F^T\mathbf{G}(\mathbf{e}(i))\mathbf{D}_S\mathbf{U}_F-\mathbf{U}_F^T\mathbf{G}(\mathbf{e}(i))\mathbf{D}_S\mathbf{U}_F\otimes\mathbf{I} \\ +\mu^2\mathbf{U}_F^T\mathbf{G}(\mathbf{e}(i))\mathbf{D}_S\mathbf{U}_F\otimes\mathbf{U}_F^T\mathbf{G}(\mathbf{e}(i))\mathbf{D}_S\mathbf{U}_F \end{bmatrix}\mathrm{vec}\left\{\mathrm{E}\left[\tilde{\mathbf{s}}_F(i)\tilde{\mathbf{s}}_F^T(i)\right]\right\} \\
&\quad +\mu^2\mathrm{vec}\left\{\mathrm{E}\left[\mathbf{U}_F^T\mathbf{G}(\mathbf{e}(i))\mathbf{D}_S\mathbf{w}(i)\mathbf{w}^T(i)\mathbf{D}_S\mathbf{G}(\mathbf{e}(i))\mathbf{U}_F\right]\right\} \\
&= \begin{bmatrix} \mathbf{I}\otimes\mathbf{I}-2\mathbf{I}\otimes\mathbf{U}_F^T\mathbf{G}(\mathbf{e}(i))\mathbf{D}_S\mathbf{U}_F \\ +\mu^2\mathbf{U}_F^T\mathbf{G}(\mathbf{e}(i))\mathbf{D}_S\mathbf{U}_F\otimes\mathbf{U}_F^T\mathbf{G}(\mathbf{e}(i))\mathbf{D}_S\mathbf{U}_F \end{bmatrix}\mathrm{vec}\left\{\mathrm{E}\left[\tilde{\mathbf{s}}_F(i)\tilde{\mathbf{s}}_F^T(i)\right]\right\} \\
&\quad +\mu^2\mathrm{vec}\left\{\mathrm{E}\left[\mathbf{U}_F^T\mathbf{G}(\mathbf{e}(i))\mathbf{D}_S\mathbf{w}(i)\mathbf{w}^T(i)\mathbf{D}_S\mathbf{G}(\mathbf{e}(i))\mathbf{U}_F\right]\right\}
\end{aligned} \tag{32}
$$

In (32), since $\mathbf{G}(\mathbf{e}(i))$ is the error processing matrix used to suppress large outliers, when a large outlier occurs, each element on the diagonal of $\mathbf{G}\left(\mathbf{e}(i)\right)$ tends to be close to 0. At the same time, the value of the step size tends to be less than 1, so that (33) is established.

$$\mu^2\mathbf{U}_F^T\mathbf{G}\left(\mathbf{e}(i)\right)\mathbf{D_S}\mathbf{U}_F\otimes\mathbf{U}_F^T\mathbf{G}\left(\mathbf{e}(i)\right)\mathbf{D_S}\mathbf{U}_F\approx\mathbf{0}^{F^2\times F^2} \tag{33}$$

Based on (32) and (33), (34) can be obtained as follows

$$\begin{aligned}&\mathrm{vec}\left\{\mathrm{E}\left[\tilde{\mathbf{s}}_F(i+1)\tilde{\mathbf{s}}_F^T(i+1)\right]\right\}\\&=\left[\mathbf{I}\otimes\mathbf{I}-2\mathbf{I}\otimes\mathbf{U}_F^T\mathbf{G}\left(\mathbf{e}(i)\right)\mathbf{D_S}\mathbf{U}_F\right]\mathrm{vec}\left\{\mathrm{E}\left[\tilde{\mathbf{s}}_F(i)\tilde{\mathbf{s}}_F^T(i)\right]\right\}\\&\quad+\mu^2\mathrm{vec}\left\{\mathrm{E}\left[\mathbf{U}_F^T\mathbf{G}\left(\mathbf{e}(i)\right)\mathbf{D_S}\mathbf{w}(i)\mathbf{w}^T(i)\mathbf{D_S}\mathbf{G}\left(\mathbf{e}(i)\right)\mathbf{U}_F\right]\right\}\end{aligned} \tag{34}$$

$\mathbf{I}\otimes\mathbf{U}_F^T\mathbf{G}\left(\mathbf{e}(i)\right)\mathbf{D_S}\mathbf{U}_F\in\mathbf{R}^{F^2\times F^2}$ is a symmetric matrix, whose spectral decomposition result is shown in (35)

$$\mathbf{I}\otimes\mathbf{U}_F^T\mathbf{G}\left(\mathbf{e}(i)\right)\mathbf{D_S}\mathbf{U}_F=\mathbf{M\Pi M}^T \tag{35}$$

where $\mathbf{\Pi}$ is a diagonal matrix consisting of the eigenvalues of $\mathbf{I}\otimes\mathbf{U}_F^T\mathbf{G}\left(\mathbf{e}(i)\right)\mathbf{D_S}\mathbf{U}_F$ . $\mathbf{M}\in\mathbf{R}^{F^2\times F^2}$ is the corresponding orthogonal matrix, which satisfies $\mathbf{MM}^T=\mathbf{M}^T\mathbf{M}=\mathbf{I}$ . Since the eigenvalues of the unitary matrix are all 1, $\mathbf{I}\otimes\mathbf{U}_F^T\mathbf{G}\left(\mathbf{e}(i)\right)\mathbf{D_S}\mathbf{U}_F$ has the same number of F reals for each eigenvalue of $\mathbf{U}_F^T\mathbf{G}\left(\mathbf{e}(i)\right)\mathbf{D_S}\mathbf{U}_F$ .

Given that $\mu^2\mathrm{vec}\left\{\mathrm{E}\left[\mathbf{U}_F^T\mathbf{G}\left(\mathbf{e}(i)\right)\mathbf{D_S}\mathbf{w}(i)\mathbf{w}^T(i)\mathbf{D_S}\mathbf{G}\left(\mathbf{e}(i)\right)\mathbf{U}_F\right]\right\}$ contains no instance of $\mathrm{E}\left[\tilde{\mathbf{s}}_F(i+1)\tilde{\mathbf{s}}_F^T(i+1)\right]$ , it follows that $\mu^2\mathrm{vec}\left\{\mathrm{E}\left[\mathbf{U}_F^T\mathbf{G}\left(\mathbf{e}(i)\right)\mathbf{D_S}\mathbf{w}(i)\mathbf{w}^T(i)\mathbf{D_S}\mathbf{G}\left(\mathbf{e}(i)\right)\mathbf{U}_F\right]\right\}$ can be regarded as a constant, to ensure that the GSP HQC algorithm remains convergence-stable in the mean-square sense, the magnitude of each component in $\mathrm{vec}\left\{\mathrm{E}\left[\tilde{\mathbf{s}}_F(i)\tilde{\mathbf{s}}_F^T(i)\right]\right\}$ at time i must be less than the magnitude of each component in $\mathrm{vec}\left\{\mathrm{E}\left[\tilde{\mathbf{s}}_F(i+1)\tilde{\mathbf{s}}_F^T(i+1)\right]\right\}$ at time (i+1).

Therefore, (36) must be satisfied

$$\left|1-2\lambda_{\max}\left(\mathbf{U}_F^T\mathbf{G}\left(\mathbf{e}(i)\right)\mathbf{D_S}\mathbf{U}_F\right)\right|<1 \tag{36}$$

According to (36), the range of step sizes that ensures that the GSP HQC algorithm remains convergence-stable in the mean-square sense can be obtained as shown in (37)

$$0<\mu<\frac{1}{\lambda_{\max}\left(\mathbf{U}_F^T\mathbf{G}\left(\mathbf{e}(i)\right)\mathbf{D_S}\mathbf{U}_F\right)} \tag{37}$$

In the frequency domain, the linearized homogeneous term of the algorithm can be written as

$$\mathbf{s}_{k+1} = \left(\mathbf{I} - c\mu\mathbf{L}\right)\mathbf{s}_{k+1} + \mathbf{O} \tag{38}$$

where $c$ is a constant, $\mathbf{O}$ is the driving term, and $\mathbf{L} = \mathbf{U}_F^T \mathbf{G} \mathbf{D}_\mathrm{S} \mathbf{U}_F$. Let $\{\lambda_i, \ \phi_i^T\}$ be the eigenpair of $\mathbf{L}$, and denote $\tilde{\mathbf{s}}_{i,k} = \phi_i^T \mathbf{s}_k$ as the coordinate of the *i*-th spectral mode; then each mode evolves independently as

$$\tilde{\mathbf{s}}_{i,k+1} = \left(\mathbf{I} - c\mu\lambda_i\right)\tilde{\mathbf{s}}_{k+1} + \tilde{r}_{i,k} \tag{39}$$

where $\tilde{r}_{i,k}$ is the driving term for the *i*-th mode, obtained by projecting the observation noise and the inhomogeneous term onto this mode. Consequently, the linear convergence factor for this mode is $\left|\mathbf{I} - c\mu\lambda_i\right|$. Hence, modes with larger eigenvalues $\lambda_i$ exhibit a smaller decay factor and converge faster; conversely, if the factor for a mode approaches zero, that mode is barely updated during the iterations, leading to extremely slow convergence. For HQC, thanks to its property of suppressing weights at large residuals, noise-dominated directions are relatively attenuated, while signal-dominated directions are relatively preserved. Therefore, under non- Gaussian noise, HQC achieves stronger suppression of high-frequency interference and faster convergence of meaningful high-frequency components by both reducing the gain for noise pseudo-modes and maintaining or enhancing the gain for true modes.

### 3.3 Steady-state error analysis

The error expression of the adaptive graph signal processing algorithm is shown in (40)

$$\mathrm{MSD_G}\left(i\right) = \mathrm{Tr}\left\{ \mathrm{E}\left[\tilde{\mathbf{s}}_F\left(i\right)\tilde{\mathbf{s}}_F^T\left(i\right)\right] \right\} \tag{40}$$

Analyzing (30) and (40), it is obvious that the steady-state error expression of the GSP HQC algorithm contains the error processing function $\mathbf{G}\left(\mathbf{e}\left(i\right)\right)$. Since $\mathbf{G}\left(\mathbf{e}\left(i\right)\right)$ exists in $1\big/\sqrt{1+\tau \mathrm{e}^2}$, to obtain the steady-state form of $\mathbf{G}\left(\mathbf{e}\left(i\right)\right)$, the Taylor expansion of the function $u\left(x\right) = 1/\sqrt{1+ax^2}$ is studied first, as shown in (41)

$$u\left(x\right) = u\left(0\right) + u'\left(0\right)x + \frac{u''\left(0\right)}{2!}x^2 + \frac{u'''\left(0\right)}{3!}x^3 + \ldots + \tag{41}$$

Based on (41), the second-order Taylor expansion of $u\left(x\right)$ can be obtained, which is shown in (42)

$$u(x) = 1 - \frac{1}{2}ax^2 \tag{42}$$

$\hat{\mathbf{s}}_F(i) \approx \mathbf{s}_F$ holds when the GSP HQC algorithm reaches steady state, thus, it can be obtained that

$$\begin{aligned} \mathbf{e}(i) &= \mathbf{D}_\mathbf{S}\left(\mathbf{U}_F\mathbf{s}_F + \mathbf{w}(i) - \mathbf{U}_F\hat{\mathbf{s}}_F(i)\right) \\ &\approx \mathbf{D}_\mathbf{S}\mathbf{w}(i) \end{aligned} \tag{43}$$

Based on (43), to obtain the steady-state error expression of the GSP HQC algorithm, start by studying the properties of $\mathbf{w}(i)$. Since both $\boldsymbol{\eta}(i)$ and $\boldsymbol{\gamma}(i)$ are Gaussian noise signals with zero mean and satisfying normal distribution, and $\delta_\gamma^2 >> \delta_\eta^2$. Therefore, (44) holds

$$\mathbf{w}(i) = \boldsymbol{\eta}(i) + \mathbf{B}\boldsymbol{\gamma}(i) = \begin{cases} (1-Pr)\mathrm{N}\left(0, \delta_\eta^2\right) \\ Pr\mathrm{N}\left(0, \delta_\eta^2 + \delta_\gamma^2\right) \approx Pr\mathrm{N}\left(0, \delta_\gamma^2\right) \end{cases} \tag{44}$$

where $\mathrm{N}(\mathrm{x}, \mathrm{y})$ denotes a normal distribution with mean x and variance y.

Since a vector with zero mean and satisfying the normal distribution divided by its standard deviation satisfies the standard normal distribution. Also, each component of a vector that satisfies the standard normal distribution also satisfies the standard normal distribution, thus

$$\begin{aligned} &\mathrm{E}\left[\ \left|\mathbf{w}(i)\right|^k\right] \\ &= Pr\mathrm{E}\left[\ \left|\boldsymbol{\gamma}(i)\right|^k\right] + (1-Pr)\mathrm{E}\left[\ \left|\boldsymbol{\eta}(i)\right|^k\right] \\ &= \left\{Pr\delta_\gamma^k\mathrm{E}\left[\left|\frac{\boldsymbol{\gamma}(i)}{\delta_\gamma}\right|^k\right] + (1-Pr)\delta_\eta^k\mathrm{E}\left[\left|\frac{\boldsymbol{\eta}(i)}{\delta_\eta}\right|^k\right]\right\} \\ &= \left[Pr\delta_\gamma^k\theta(k) + (1-Pr)\delta_\eta^k\theta(k)\right]\ \boldsymbol{\upsilon} \end{aligned} \tag{45}$$

where $\theta(k)$ is the kth order moment of origin of the absolute value of the variable whose probability density function satisfies the standard normal distribution, and when k is determined, the value of $\theta(k)$ is subsequently determined. $\boldsymbol{\upsilon} \in \mathbf{R}^{N\times 1}$ is a vector with all components 1.

Based on (30), (42), and (45), the Taylor expansion of $\mathbf{G}(\mathbf{e}(i))$ at $\mathbf{e}(i) = \mathbf{0}$ under steady-state conditions can be obtained as

$$\mathbf{G}(\mathbf{e}(i)) = \left[1 - \frac{\tau}{2}Pr\delta_\gamma^2\theta(2) - \frac{\tau}{2}(1-Pr)\delta_\eta^2\theta(2)\right]\mathbf{D}_\mathbf{S}\mathbf{I} \tag{46}$$

The validity of this second-order approximation statistically requires that the second-order moment is sufficiently small, i.e., $\tau \max_{1\le j\le N} \mathrm{E}\left[e_j^2(i)\right]$.

Taking the squares of the weighted Euclidean parameter on both sides of (30), one can obtain that

$$\begin{aligned} \mathrm{E}\left[\ \left\|\tilde{\mathbf{s}}_F(i+1)\right\|_{\mathbf{K}}^2\ \right] &= \mathrm{E}\left[\ \left\|\tilde{\mathbf{s}}_F(i+1)\right\|_{\mathbf{T}}^2\ \right] \\ &+ \mu^2 \mathrm{Tr}\left[\mathbf{K}\mathbf{U}_F^T\mathbf{G}(\mathbf{e}(i))\mathbf{D}_\mathbf{S}\mathbf{w}(i)\mathbf{w}^T(i)\mathbf{D}_\mathbf{S}\mathbf{G}(\mathbf{e}(i))\mathbf{U}_F\right] \end{aligned} \tag{47}$$

where $\mathrm{Tr}(\cdot)$ denotes the trace operation, $\mathbf{K}\in\mathbf{R}^{F\times F}$ is a freely selectable nonnegative fixed weight matrix. The relationship satisfied between $\mathbf{K}$ and $\mathbf{T}\in\mathbf{R}^{F\times F}$ is shown in (48)

$$\mathbf{T} = \left[\mathbf{I}-\mu\mathbf{U}_F^T\mathbf{G}(\mathbf{e}(i))\mathbf{D}_\mathbf{S}\mathbf{U}_F\right]^T \mathbf{K}\left[\mathbf{I}-\mu\mathbf{U}_F^T\mathbf{G}(\mathbf{e}(i))\mathbf{D}_\mathbf{S}\mathbf{U}_F\right] \tag{48}$$

Vectorizing $\mathbf{K}$ and $\mathbf{T}$ with $\mathrm{vec}(\cdot)$, the results are shown in (49) and (50), respectively.

$$\mathbf{k} = \mathrm{vec}(\mathbf{K}) \tag{49}$$

$$\mathbf{t} = \mathrm{vec}(\mathbf{T}) \tag{50}$$

By using $\mathrm{Tr}\{\mathbf{XY}\} = \mathrm{vec}\{\mathbf{Y}^T\}^T \mathrm{vec}\{\mathbf{X}\}$ and $\mathrm{Tr}\{\mathbf{ABC}\} = (\mathbf{C}^T\otimes\mathbf{A})\mathrm{vec}\{\mathbf{B}\}$, (47) can be simplified to (51)

$$\mathrm{E}\left[\ \left\|\tilde{\mathbf{s}}_F(i+1)\right\|_{\mathbf{k}}^2\ \right] = \mathrm{E}\left[\ \left\|\tilde{\mathbf{s}}_F(i+1)\right\|_{\mathbf{Rk}}^2\right] + \mu^2 \mathrm{vec}(\mathbf{H})^T\mathbf{k} \tag{51}$$

where

$$\begin{aligned} \mathbf{H} &= \mathbf{U}_F^T\mathbf{G}(\mathbf{e}(i))\mathbf{D}_\mathbf{S}\mathbf{w}(i)\mathbf{w}^T(i)\mathbf{D}_\mathbf{S}\mathbf{G}(\mathbf{e}(i))\mathbf{U}_F \\ &= \delta_\eta^2\mathbf{U}_F^T\mathbf{G}(\mathbf{e}(i))\mathbf{D}_\mathbf{S}\mathbf{G}(\mathbf{e}(i))\mathbf{U}_F + Pr\delta_\gamma^2\mathbf{U}_F^T\mathbf{G}(\mathbf{e}(i))\mathbf{D}_\mathbf{S}\mathbf{G}(\mathbf{e}(i))\mathbf{U}_F \end{aligned} \tag{52}$$

$$\begin{cases} \mathbf{t} = \mathrm{vec}(\mathbf{T}) = \mathbf{Rk} \\ \mathbf{R} = \left[\mathbf{I}-\mu\mathbf{U}_F^T\mathbf{G}(\mathbf{e}(i))\mathbf{D}_\mathbf{S}\mathbf{U}_F\right]^T \otimes \left[\mathbf{I}-\mu\mathbf{U}_F^T\mathbf{G}(\mathbf{e}(i))\mathbf{D}_\mathbf{S}\mathbf{U}_F\right] \end{cases} \tag{53}$$

When the GSP HQC algorithm reaches steady state, it can be obtained that

$$\mathrm{E}\left[\ \left\|\tilde{\mathbf{s}}_F(i+1)\right\|^2\right] \approx \mathrm{E}\left[\ \left\|\tilde{\mathbf{s}}_F(i)\right\|^2\right] \tag{54}$$

Based on (54), when $i\to\infty$, one can obtain that

$$\mathrm{E}\left[\ \left\|\tilde{\mathbf{s}}_F(\infty)\right\|_{(\mathbf{I}-\mathbf{R})\mathbf{k}}^2\ \right] = \mu^2\mathrm{vec}(\mathbf{H})^T\mathbf{k} \tag{55}$$

By making $\mathbf{k} = (\mathbf{I}-\mathbf{R})^{-1}\mathrm{vec}(\mathbf{I})$ in equation (55), the steady state error of the GSP HQC algorithm can be obtained as shown in (56)

$$\mathrm{MSD_G}(\infty) = \mathrm{E}\left[\ \left\|\tilde{\mathbf{s}}_F(\infty)\right\|^2\right] = \mu^2\mathrm{vec}^T(\mathbf{H})(\mathbf{I}-\mathbf{R})^{-1}\mathrm{vec}(\mathbf{I}) \tag{56}$$

## 4. Computational complexity analysis

This section analyzes and compares the computational complexity of the proposed GSP HQC algorithm and other existing algorithms to complete each iteration, including addition/subtraction, multiplication/division, power multiplication, exponentiation, root sign and Direct Matrix Inversion (DMI). It should be noted that the computational complexity analyses performed in this section relate only to the updated part of the graph signal adaptive estimation formulation of each algorithm, which for the GSP HQC algorithm is $\mathbf{U}_F\mathbf{U}_F^T\mathbf{G}\left(\mathbf{e}(i)\right)\mathbf{e}(i)$, and so on.

Table 1 Analysis of the actual computational complexity of the related operations

| Operation | Multiplication | Addition |
|---|---|---|
| $\mathbf{U}_F^T\mathbf{e}(i)$ | $F\lvert\mathbf{S}\rvert$ | $F\left(\lvert\mathbf{S}\rvert-1\right)$ |
| $\mathbf{D}_\mathbf{S}\mathbf{U}_F$ | $F\lvert\mathbf{S}\rvert$ | 0 |
| $\mathbf{U}_F^T\mathbf{D}_\mathbf{S}\mathbf{U}_F$ | $F^2\lvert\mathbf{S}\rvert$ | $F^2\left(\lvert\mathbf{S}\rvert-1\right)$ |
| $\mathbf{e}^T(i)\ \mathbf{e}(i)$ | $\lvert\mathbf{S}\rvert$ | $\lvert\mathbf{S}\rvert-1$ |
| $\left(\mathbf{U}_F^T\mathbf{D}_\mathbf{S}\mathbf{U}_F\right)^{-1}$ | $\frac{1}{3}F^3$ | 0 |

Table 2 Computational complexity analysis and comparison of the related algorithms

| Algorithm | Multiplication/Division | Addition | square root | exp | power product | DMI |
|---|---|---|---|---|---|---|
| GSP LMS | $F\left(\lvert\mathbf{S}\rvert+N\right)$ | $F\lvert\mathbf{S}\rvert-F+NF-N$ | 0 | 0 | 0 | 0 |
| GSP NLMS | $\left(\lvert\mathbf{S}\rvert+N\right)\left(F^2+F\right)$ | $F^2\left(\lvert\mathbf{S}\rvert+N-1\right)+F\left(\lvert\mathbf{S}\rvert-1\right)-N$ | 0 | 0 | 0 | $\frac{1}{3}F^3$ |
| GSP MCC | $N^2(F+1)+N+\lvert\mathbf{S}\rvert$ | $N(NF-1)$ | 0 | $N$ | $N$ | 0 |
| GSP GMCC | $N^2(F+1)+2N+\lvert\mathbf{S}\rvert$ | $N(NF-1)$ | 0 | $N$ | $2N$ | 0 |
| GSP LOG | $N^2(F+1)+2N+\lvert\mathbf{S}\rvert$ | $N^2F$ | 0 | 0 | $N$ | 0 |
| GSP HQC | $N^2(F+1)+2N+\lvert\mathbf{S}\rvert$ | $N^2F$ | $N$ | 0 | $N$ | 0 |

Next, based on **Note 1** and **Note 2**, the computational complexity analysis of the related algorithms including the GSP HQC algorithm will be performed.

**Note 1:** When a matrix of shape $T_1\times T_2$ is multiplied with a matrix of shape $T_2\times T_3$, the computations in both multiplication and addition are $T_1\times T_2\times T_3$ and $T_1\times T_3\times\left(T_2-1\right)$, respectively.

**Note 2:** Some actual computational complexity of the related operations is listed in Table 1.

The following conclusions can be drawn from Table 2: the overall computational complexity of the proposed GSP HQC algorithm is N more than that of the GSP LOG algorithm, i.e., the square root section. In addition, the overall computational complexity of the GSP HQC algorithm is slightly higher than that of the GSP MCC algorithm and comparable to that of the GSP GMCC algorithm. Besides, among all the algorithms, the GSP NLMS algorithm has the highest computational complexity of all the algorithms due to the fact that the normalization part involves matrix inversion.

## 5. Simulation

Based on the July temperature data from the Brazilian meteorological temperature dataset [37], which is shown in Fig. 5, the performance of the proposed GSP HQC algorithm is validated. The related parameter settings are shown as follows: $N = 299, K = 8, |\mathbf{S}| = 210, F = 200$. Interpretation of the above parameters is: there are a total of 299 sites in this data, each site is connected to the nearest 8 sites in its vicinity, and $|\mathbf{S}| = 210$ means that there are a total of 210 sites whose data can be captured. The weight $a_{nm}$ between node $n$ and node $m$ in the graph is computed via

$$a_{nm} = \exp\left(-\frac{d_h(n,m)^2}{2\theta^2}\right) \tag{57}$$

where $d_h(n,m)$ is the Haversine distance between the two stations, and $\theta$ is the scale parameter.

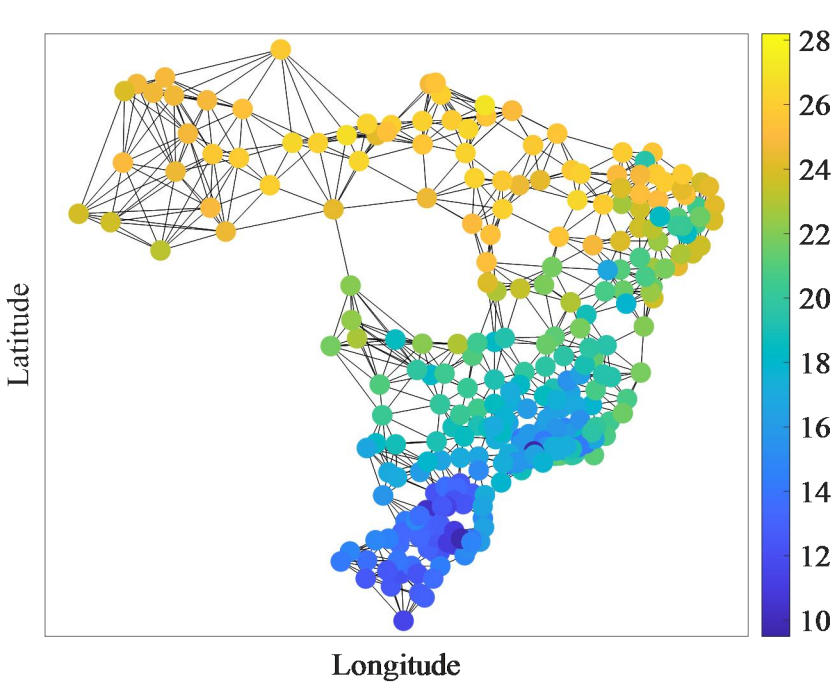

Fig. 5. July data of the Brazilian temperature dataset

In addition, the parameter $\mathrm{MSD_G}$ is used to quantitatively evaluate the performance of the relevant algorithms in terms of convergence/tracking rate and steady state error, which is shown in (58)

$$\mathrm{MSD_G} = 10\log_{10}\left\|\hat{\mathbf{x}}_{\mathbf{o}}(i) - \mathbf{x}_{\mathbf{o}}\right\|_2^2 \tag{58}$$

The random number generation process for each independent repetition of the experiments in this section is realized by using the random number seed rng(seed) in MATLAB R2024b. The value of the seed is the number of times that the experiment has been independently repeated, and the number of independent repetitions of the experiment will be determined when the seed number seed is determined, the random number will be determined with it. The simulation results of each experiment in the experimental part of the computer simulation are obtained by taking the average of 100 independent repetitions of the experiment.

## 5.1 Parameters discussion

In this section, Experiment 1 is conducted to discuss the effect of different parameters on the performance of the GSP HQC algorithm, and the relevant parameter settings for Experiment 1 are shown in Table 3. The simulation result of experiment 1 is shown in Fig. 6.

Table 3 Parameter settings related to experiment 1

| Experiment number | The related parameters | noise environment |
|---|---|---|
| Experiment 1-1 | $\mu = 0.98$<br>$\tau \in \{0.000001,\ 0.00001,\ 0.0001,\ 0.001,\ 0.01,\ 0.1,\ 1,\ 2,\ 4,\ 8\}$ | $Pr = 0.1$<br>$\delta_\eta^2 = 0.01,\ \delta_\gamma^2 = 10^6 \delta_\eta^2$ |
| Experiment 1-2 | $\mu \in \{0.000001,\ 0.00001,\ 0.0001,\ 0.001,\ 0.01,\ 0.1,\ 1,\ 2,\ 5,\ 10\}$<br>$\tau \in \{0.000001,\ 0.00001,\ 0.0001,\ 0.001,\ 0.01,\ 0.1,\ 1,\ 5,\ 10\}$ | $Pr = 0.1$<br>$\delta_\eta^2 = 0.01,\ \delta_\gamma^2 = 10^6 \delta_\eta^2$ |

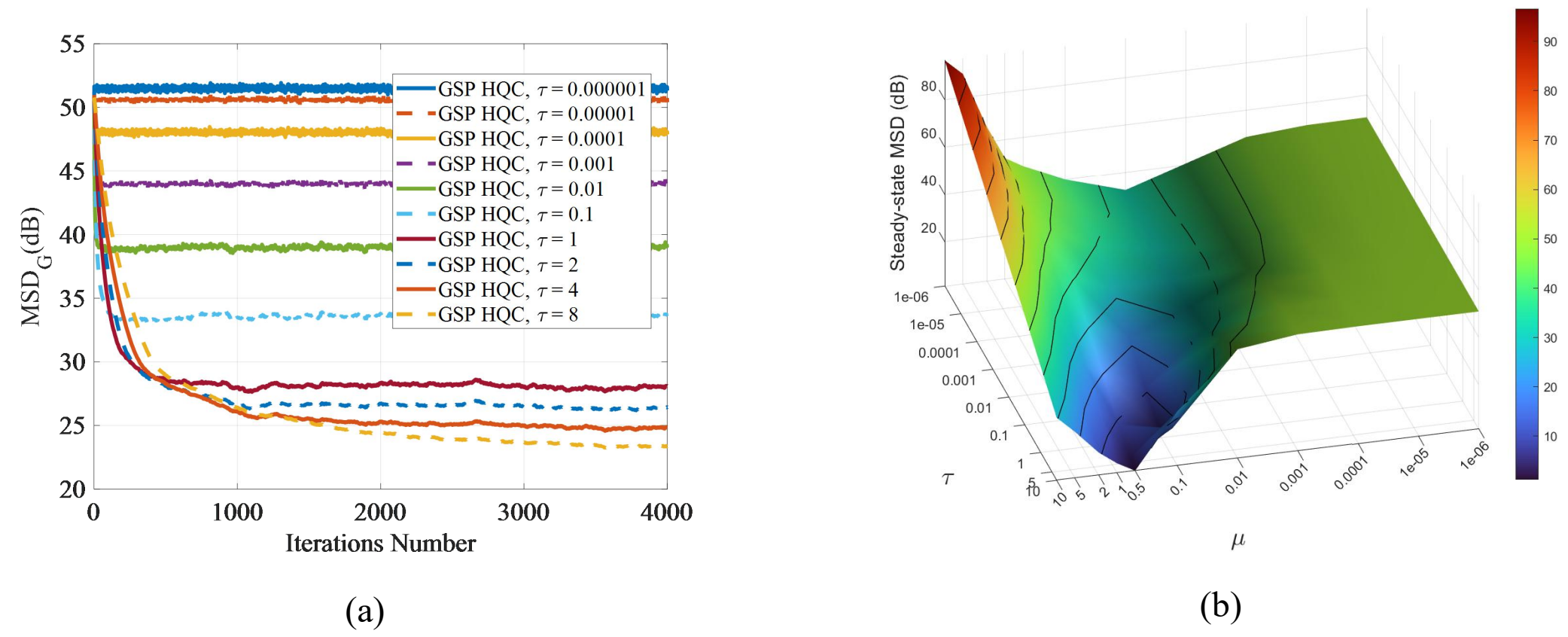

Fig. 6. The effect of different parameters on the performance of the GSP HQC algorithm
(a) Effects of Different $\tau$ Values (b) 3D Surface of Steady-State MSD versus $\tau$ and $\mu$

The following conclusions can be drawn from Fig. 6 (a): the GSP HQC algorithm with different $\tau$ shows different performance when the step size parameter and the impulse noise environment parameter are set the same. At the same time, the change of parameter $\tau$ has less influence on the convergence rate of the GSP HQC algorithm, and the curves of the corresponding GSP HQC algorithms almost coincide when $\tau \in \{1,\ 2,\ 4,\ 8\}$ . As the parameter increases, the convergence speed of the GSP HQC algorithm slows down and the steady state error increases.

Fig. 6 (b) visually illustrates the parameter coupling and the unstable region. Based on the experimental results, in order to balance the performance metrics of both convergence rate and steady state error, after discussion, the parameterization of the GSP HQC algorithm in this section is determined as follows: $\tau = 2$ , and the result of the parameterization discussion will be used in the later sections on the validation of the range of the step size constraints and the comparison of the performance of multiple algorithms. Meanwhile, when comparing the performance of multiple algorithms, a uniform step size of $\mu = 0.98$ was employed for all algorithms.

## 5.2 Validation of step size constraint range

The step size constraint range obtained in Section 4.2 is verified in this section, and the step size range is shown in (59)

$$0 < \mu < \frac{1}{\lambda_{\max}\left(\mathbf{U}_F^T \mathbf{G}\left(\mathbf{e}(i)\right)\mathbf{D}_\mathrm{S}\mathbf{U}_F\right)} \tag{59}$$

After analyzing (59), the following conclusions can be drawn:

When i = 0, each component of $\mathbf{e}(i)$ has a maximum value, and due to the presence of $1/\sqrt{1+\tau \mathrm{e}^2}$ in the error processing function $\mathbf{G}\left(\mathbf{e}(i)\right)$, $\lambda_{\max}\left(\mathbf{U}_F^T \mathbf{G}\left(\mathbf{e}(i)\right)\mathbf{D}_\mathrm{S}\mathbf{U}_F\right)$ is the smallest at this time, i.e., at this time, the step range represented by (59) is the largest.

When $i \to \infty$ , each component of $\mathbf{e}(i)$ has a minimum value, and due to the presence of $1/\sqrt{1+\tau \mathrm{e}^2}$ in the error processing function $\mathbf{G}\left(\mathbf{e}(i)\right)$, $\lambda_{\max}\left(\mathbf{U}_F^T \mathbf{G}\left(\mathbf{e}(i)\right)\mathbf{D}_\mathrm{S}\mathbf{U}_F\right)$ is the biggest at this time, i.e., at this time, the step range represented by (59) is the smallest and most accurate.

It should be noted that this result remains a conservative approximation, as Assumption 3 and Eq. (33) used in the derivation are crude approximations.

Based on the above analysis, Experiment 2 and Experiment 3 are conducted in this section to verify the correctness of (59) when it tends to be stabilized, and the related parameters of the experiments are set as shown in Table 4. When the second half of the iteration starts, Experiment 2 and Experiment 3 decompose the eigenvalue of $\mathbf{U}_F^T \mathbf{G}\left(\mathbf{e}(i)\right)\mathbf{D}_S \mathbf{U}_F$ at each moment and determine the critical value of the step size according to (59), and the correctness of (59) is verified by comparing the steady state error of GSP HQC algorithms corresponding to different multiples of the critical value with that of the first half of the iteration.

Table 4 Parameter settings related to Experiment 2 and Experiment 3

| **Experiment number** | **The related parameters** | **Noise environment** |
|---|---|---|
| Experiment 2 | Initial step size: $\mu = 0.8,\ \tau = 2$,<br>iterationnumber: 4000<br>critical value multiple: $k \in \{0.4,\ 0.8,\ 1.0,\ 1.2,\ 1.6\}$ | $Pr = 0.1$<br>$\delta_\eta^2 = 0.01,\ \delta_\gamma^2 = 10^6 \delta_\eta^2$ |
| Experiment 3 | Initial step size: $\mu = 0.8,\ \tau = 2$,<br>iterationnumber: 4000<br>critical value multiple: $k \in \{0.4,\ 0.8,\ 1.0,\ 1.2,\ 1.6\}$ | $Pr = 0.15$<br>$\delta_\eta^2 = 0.01, \delta_\gamma^2 = 10^6 \delta_\eta^2$ |

The simulation results of Experiment 2 and Experiment 3 are shown in Fig. 7 and Fig. 8, respectively.

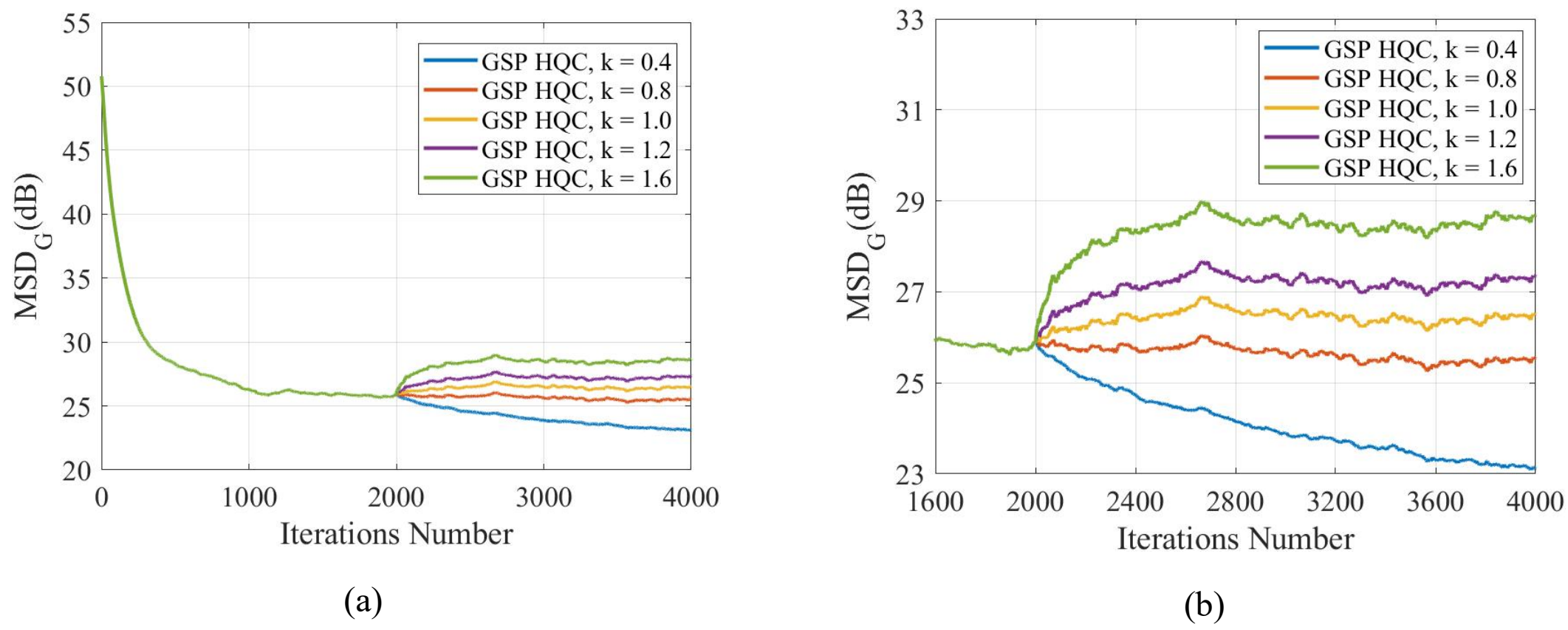

Fig. 7. Validation of the step-size constraint range of the GSP HQC algorithm when $Pr = 0.1$

(a) overall situation (b) latter part

After analyzing Fig. 7, the following conclusions can be drawn:

When $k \in \{0.4,\ 0.8\}$, the GSP HQC algorithm is in a state of convergence compared to the GSP HQC algorithm when $k = 1.0$.

When $k \in \{1.2,\ 1.6\}$, the GSP HQC algorithm is in a divergent state compared to the GSP HQC algorithm when $k = 1.0$, and the divergent state of the GSP HQC algorithm intensifies as the value of k increases.

When $k=0.8$, the steady state error of the GSP HQC algorithm in the second half of the iteration is nearly the same as that in the first half of the iteration, while the steady state error of the GSP HQC algorithm of $k=1.0$ in the second half of the iteration is slightly larger than that in the first half of the iteration, but according to (59) the GSP HQC algorithm of $k=1.0$ is already in the divergence state. Therefore, the correctness of (59) is verified.

Analyzing Fig. 8 leads to similar conclusions as analyzing Fig. 7.

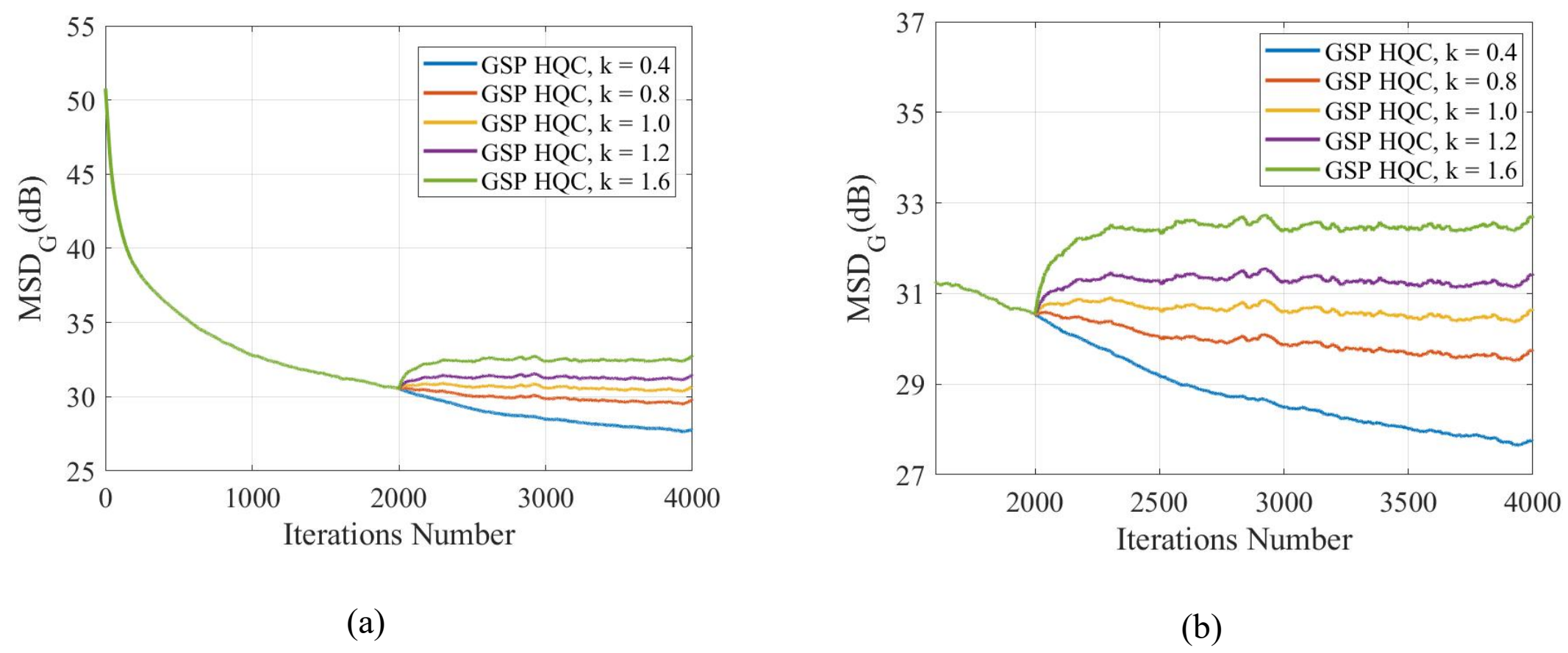

Fig. 8. Validation of the step-size constraint range of the GSP HQC algorithm when $Pr=0.15$

(a) overall situation (b) latter part

## 5.3 Algorithms performance comparison

In addition to considering the common impulse noise with clear and well-defined mathematical properties, this section also considers the Cauchy noise and Laplace noise of $\alpha$ stable distribution noise, whose functions are shown in (60) and (61) respectively [47].

$$\phi(k)=\exp\left\{j\delta k-\gamma|k|^{\alpha}\left[1+j\beta\operatorname{sgn}(k)S(k,\alpha)\right]\right\} \tag{60}$$

$$S(k,\alpha)=\begin{cases}\tan\left(\dfrac{\alpha\pi}{2}\right), & \text{if } \alpha\neq 1\\ \dfrac{2}{\pi}\log|k| , & \text{if } \alpha=1\end{cases} \tag{61}$$

When $\alpha=1,\ \delta=0$, the $\alpha$ stable distribution noise will evolve into Cauchy noise. $\beta$ is the skewness parameter, and $\gamma$ is the scale parameter.

The probability density function of Laplace noise is shown in (62)

$$f(x\mid\mu,b)=\frac{1}{2b}\exp\left(-\frac{|x-\mu|}{b}\right) \tag{62}$$

where $\mu$ is the mean parameter and $b$ is the position parameter. Laplace noise is considered to occur when $\mu$ and $b$ take values of 0 and 1, respectively.

When deriving the GSP HQC algorithm, it has been mentioned that: there is only a difference between the results derived by the LOG function and the HQC function in the presence or absence of a root sign. Therefore, the error handling function $\mathbf{F}(\mathbf{e}(i))$ of the GSP LOG algorithm is shown in (63)

$$\mathbf{F}(\mathbf{e}(i))=\begin{bmatrix} \frac{1}{1+\alpha e_1^2(i)} & 0 & 0 & \cdots & 0 \\ 0 & \frac{1}{1+\alpha e_2^2(i)} & 0 & 0 & \cdots \\ 0 & 0 & \frac{1}{1+\alpha e_3^2(i)} & & \cdots \\ \vdots & \vdots & \vdots & & \cdots \\ 0 & 0 & \cdots & & \frac{1}{1+\alpha e_N^2(i)} \end{bmatrix} \tag{63}$$

In the following, the Experiment 4 to Experiment 7 are conducted to investigate the performance advantages of the GSP HQC algorithm over other algorithms in different non-Gaussian noise environments, with the parameter settings in the impulse noise environment as shown in Table 5,The algorithm parameters are identical across these experiments. The light-colored shaded bands in Fig. 9 represent the fluctuation range of the learning curves across multiple independent repeated trials: For each algorithm, we conducted 100 independent trials. At the *i*-th iteration, we recorded the mean-square error in the linear domain, $\mathrm{MSD}_r(i)(r=1,\ldots,100)$, for each trial, and then calculated the sample mean

$$\overline{\mathrm{MSD}}(i)=\frac{1}{100}\sum_{i}^{100}\mathrm{MSD}_r(i) \tag{64}$$

and sample standard deviation

$$\sigma_{\mathrm{MSD}(i)}=\sqrt{\frac{1}{100-1}\sum_{i}^{100}\left(\mathrm{MSD}_r(i)-\overline{\mathrm{MSD}}(i)\right)^2} \tag{65}$$

To avoid introducing bias from the dB transformation, the upper and lower bounds of the shaded band were first computed in the linear domain as $\overline{\mathrm{MSD}}\pm\sigma_{\mathrm{MSD}}$, and then converted to dB.

Table 5 The related settings about Experiment 4 to Experiment 7

| Experiment number | Algorithm | The related parameters | noise environment |
|---|---|---|---|
| Experiment 4 | GSP LMS | $\mu = 0.98$ | BG noise<br>$Pr = 0.1$<br>$\delta_\eta^2 = 0.01,\ \delta_\gamma^2 = 10^6 \delta_\eta^2$ |
| Experiment 5 | GSP NLMS | $\mu = 0.07$ | BG noise<br>$Pr = 0.15$<br>$\delta_\eta^2 = 0.01,\ \delta_\gamma^2 = 10^6 \delta_\eta^2$ |
| Experiment 6 | GSP MCC<br>GSP GMCC | $\mu = 0.98,\ \lambda = 0.01$<br>$\mu = 0.98,\ \alpha = 1.4,\ \lambda = 0.01$ | Cauchy noise<br>$\beta = 0, \gamma = 1$ |
| Experiment 7 | GSP LOG<br>GSP HQC | $\mu = 0.98,\ \alpha = 0.06$<br>$\mu = 0.98,\ \tau = 2$ | Laplacian noise<br>$\mu = 0, b = 1$ |

where the noise environment set up in Experiment 6 is the Cauchy noise environment, the noise environment set up in Experiment 7 is the Laplace noise environment, and the parameters of the relevant algorithms are exactly the same as in Table 5. The basis of the related parameter settings in Table 5 is to ensure that the GSP MCC algorithm, the GSP GMCC algorithm, the GSP LOG algorithm, and the GSP HQC algorithm are able to maintain approximately the same convergence/tracking speed in different non-Gaussian noise environments. Meanwhile, each experiment was set to multiply the value of the original graph signal by a factor of 1.4 starting from the second half of the iteration to simulate the abrupt change, which is used to test the tracking rate of the related algorithms. The simulation results of Experiment 4, Experiment 5 are shown in Fig. 9(a) and Fig. 9(b), respectively.

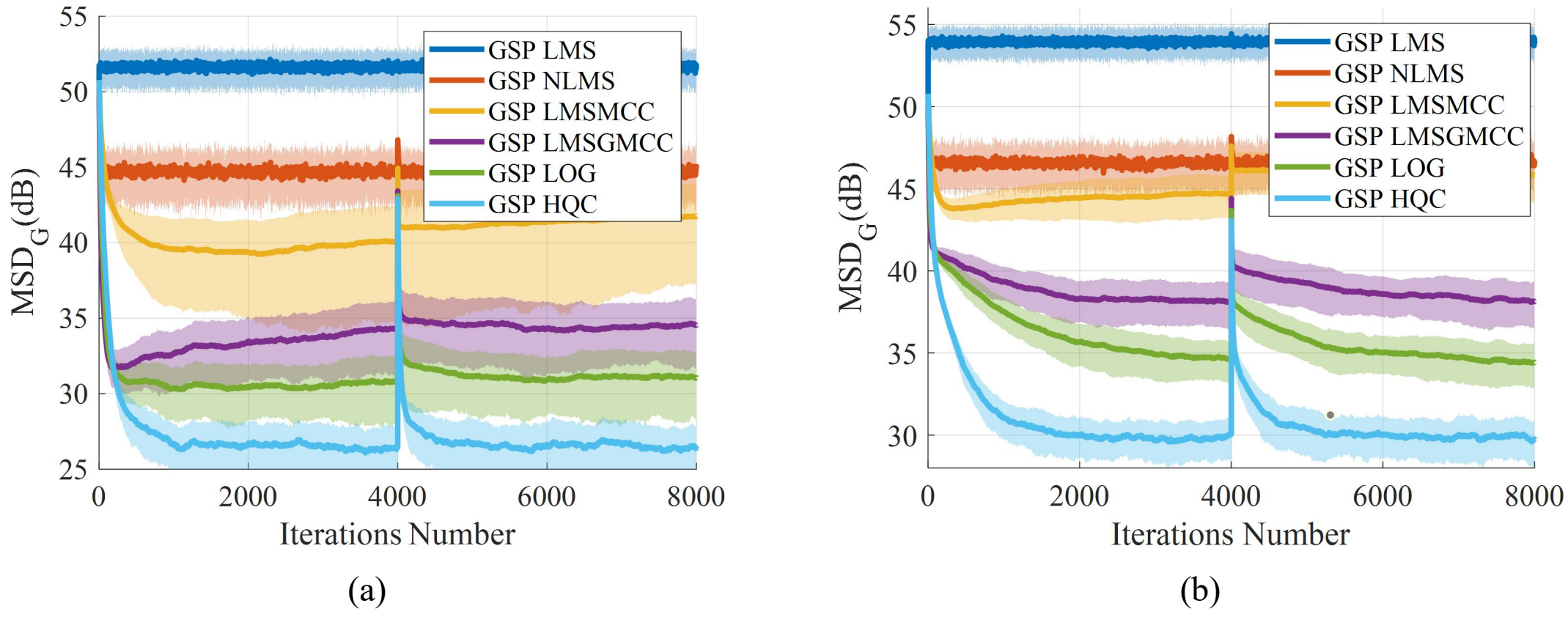

Fig. 9 Comparison of the performance of each algorithm in impulse noise environment (a) $Pr = 0.1$ (b) $Pr = 0.15$

The simulation result of Experiment 6 is shown in Fig. 10.

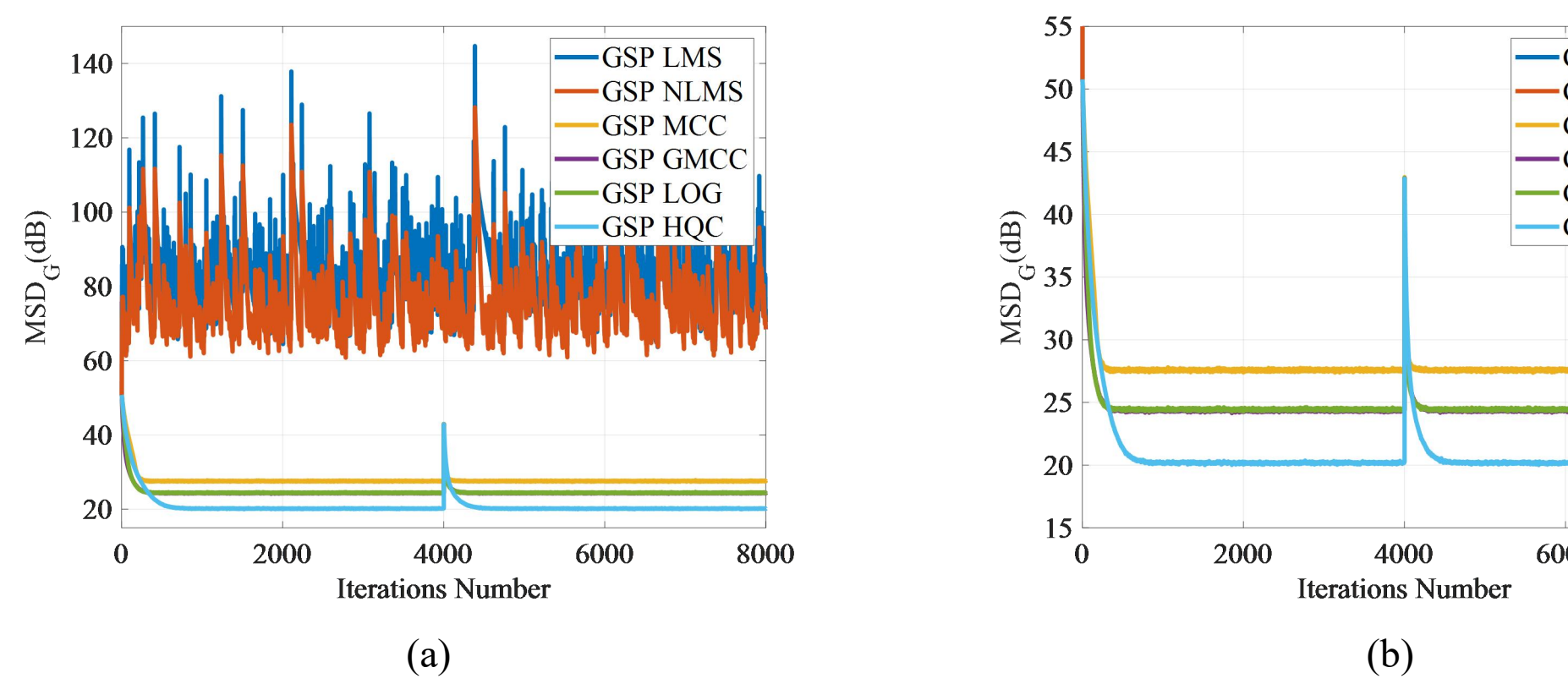

Fig. 10 Comparison of the performance of each algorithm in Cauchy noise environment (a) Overall (b) Local

The simulation results of Experiment 7 are shown in Fig. 11.

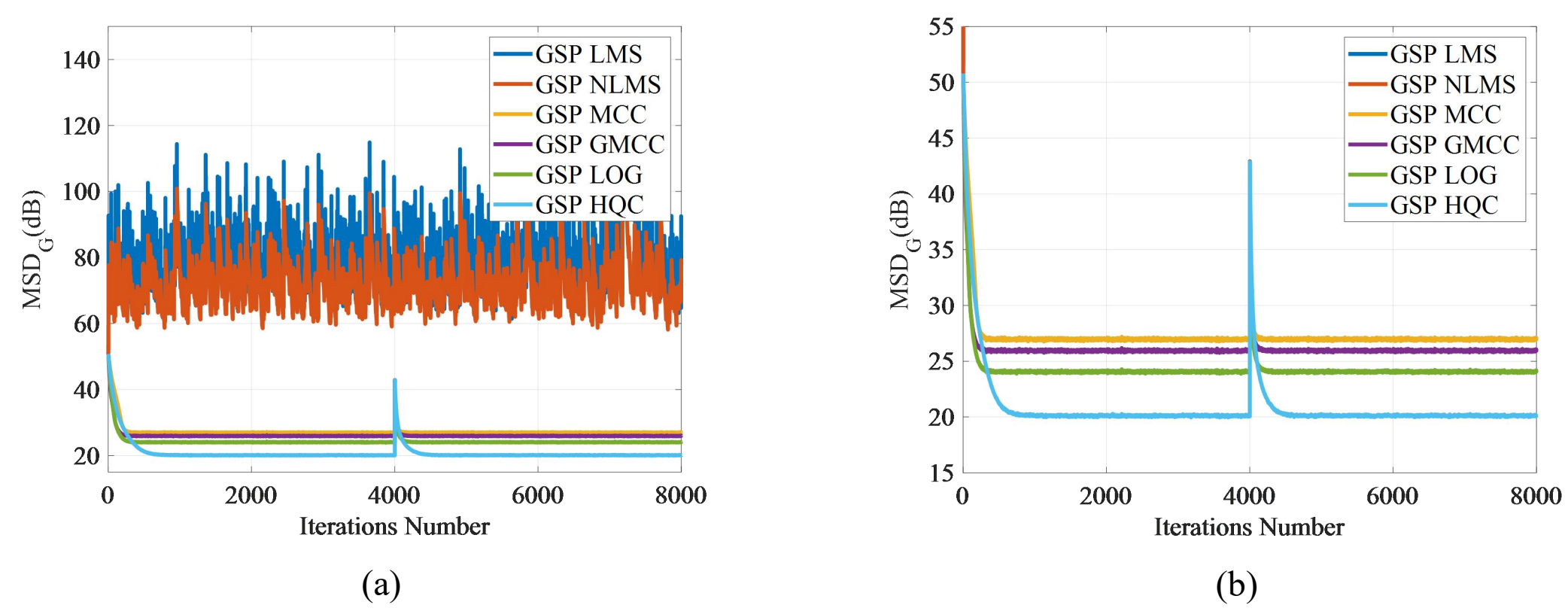

Fig. 11 Comparison of the performance of each algorithm in Laplace noise environment (a) Overall (b) Local

From Table 5, Fig. 9, Fig, 10 and Fig. 11, it can be found that:

While ensuring nearly identical convergence/tracking rate, the GSP HQC algorithm has the lowest steady-state error in different non-Gaussian noise environments, which indicates that its adaptive estimation accuracy is the highest.

While the overall computational complexity is comparable, GSP HQC requires fewer parameters to be debugged and has higher adaptive estimation accuracy than the GSP GMCC algorithm.

Compared with the impulse noise, the fluctuation degree of the $\mathrm{MSD}_G$ curves of the GSP LMS algorithm and the GSP NLMS algorithm in Cauchy noise and Laplace noise environment is more drastic, which indicates that the deterioration of the performance of the GSP LMS algorithm and the GSP NLMS algorithm in Cauchy noise and Laplace noise environments is further aggravated. Meanwhile, algorithms such as the GSP GMCC algorithm and the GSP HQC show better robustness.

Next, the Experiment 8 and Experiment 9 are conducted for the comparison of convergence rate. The relevant algorithm parameters and noise environment settings are shown in Table 6. The relevant parameters are set on the basis of ensuring that each algorithm arrives at a near-uniform steady-state error. The related simulation results are shown in Fig. 12(a) and Fig. 12(b), respectively.

Table 6 Parameter settings related to Experiment 8 and Experiment 9

| **Experiment number** | **Algorithm** | **The related parameters** | **noise environment** |
|---|---|---|---|
| Experiment 8 | GSP LOG | $\mu = 0.7,\ \alpha \in \{1, 2, 3\}$ | $Pr = 0.05$<br>$\delta_\eta^2 = 0.01,\ \delta_\gamma^2 = 10^6 \delta_\eta^2$ |
| | GSP HQC | $\mu = 0.98,\ \tau = 0.5$ | |
| | GSP GMCC | $\mu = 0.05,\ \lambda = 0.01, \alpha = 1.4$ | |
| Experiment 9 | GSP LOG | $\mu = 0.98,\ \alpha \in \{1, 2, 3\}$ | $Pr = 0.10$<br>$\delta_\eta^2 = 0.01,\ \delta_\gamma^2 = 10^6 \delta_\eta^2$ |
| | GSP HQC | $\mu = 0.98,\ \tau = 0.5$ | |
| | GSP GMCC | $\mu = 0.1,\ \lambda = 0.01, \alpha = 1.4$ | |

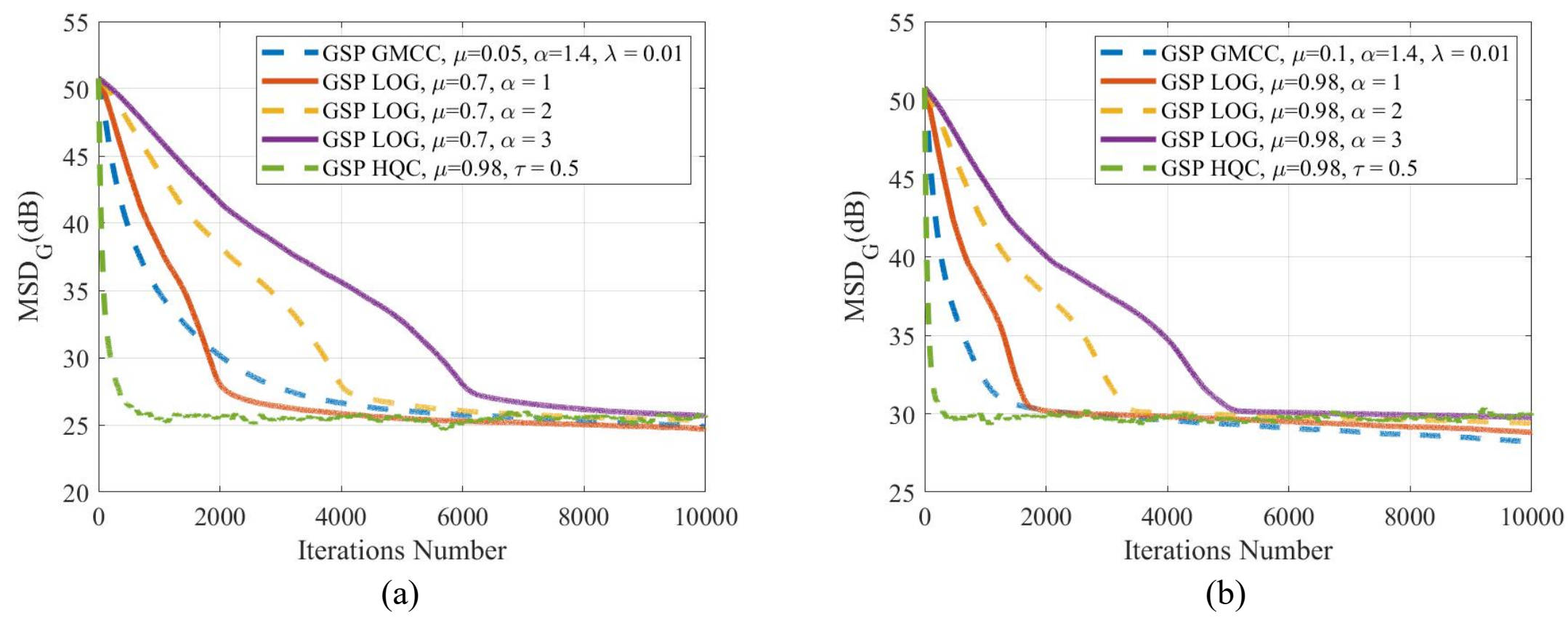

Fig. 12 Comparison of convergence rate between GSP HQC algorithm and GSP LOG algorithm

(a) $Pr = 0.05$ (b) $Pr = 0.10$

To address the phenomenon of unstable error fluctuation after the algorithms reach the steady state, in [36], the authors take the value of 1.03 times the steady state error value of the algorithms as a reference value and quantitatively analyze the superior performance of their proposed algorithms in terms of convergence speed by counting the number of iterations that the error value of each algorithm has gone through to reach the reference value in an iteration. In Fig. 12(a) and Fig. 12(b), the steady state errors of each algorithm are 25 dB and 30 dB, respectively. Based on the methodology adopted in [36], this section counts the number of iterations in which the error value of each algorithm in Fig. 12(a) and Fig. 12(b) reaches the reference value for the first time in this section, and the statistical results are shown in the tables as shown in Table 7 and Table 8, respectively.

Table 7 Quantitative statistical analysis of the convergence rate of the related algorithms in Experiment 8

| Experiment number | Algorithm and corresponding parameters | Number of iterations after which the error is less than the reference value for the first time (times) | | |
|---|---|---|---|---|
| | | 35 dB | 30 dB | 25.75 dB |
| Experiment 8 | GSP LOG, $\mu = 0.7, \alpha = 1$ | 1408 | 1826 | 4208 |
| | GSP LOG, $\mu = 0.7, \alpha = 2$ | 2827 | 3729 | 7053 |
| | GSP LOG, $\mu = 0.7, \alpha = 3$ | 4229 | 5629 | 9632 |
| | GSP HQC, $\mu = 0.98, \tau = 0.5$ | 84 | 199 | 687 |
| | GSP GMCC, $\mu = 0.05,\ \lambda = 0.01,\ \alpha = 1.4$ | 976 | 2042 | 5924 |

Table 8 Quantitative statistical analysis of the convergence rate of the related algorithms in Experiment 9

| Experiment number | Algorithm and corresponding parameters | Number of iterations after which the error is less than the reference value for the first time (times) | | |
|---|---|---|---|---|
| | | 40 dB | 35 dB | 30.9 dB |
| Experiment 9 | GSP LOG, $\mu = 0.98, \alpha = 1$ | 673 | 1300 | 1646 |
| | GSP LOG, $\mu = 0.98, \alpha = 2$ | 1344 | 2612 | 3225 |
| | GSP LOG, $\mu = 0.98, \alpha = 3$ | 2015 | 3934 | 4825 |
| | GSP HQC, $\mu = 0.98, \tau = 0.5$ | 35 | 74 | 185 |
| | GSP GMCC, $\mu = 0.1,\ \lambda = 0.01,\ \alpha = 1.4$ | 271 | 632 | 1301 |

## 5.4 Temperature prediction applications

The U.S. weather temperature dataset is a time-varying streaming graph signal, as shown in Fig. 13 [30] The related parameters are set as follows: $K = 7$, $N = 205$, $|\mathbf{S}| = 130$, $F = 125$. The weights of the graph are also computed via Eq. (57). The noise environment settings in this section are identical to 5.3. The related algorithms and their corresponding parameters can be found in Table 9.

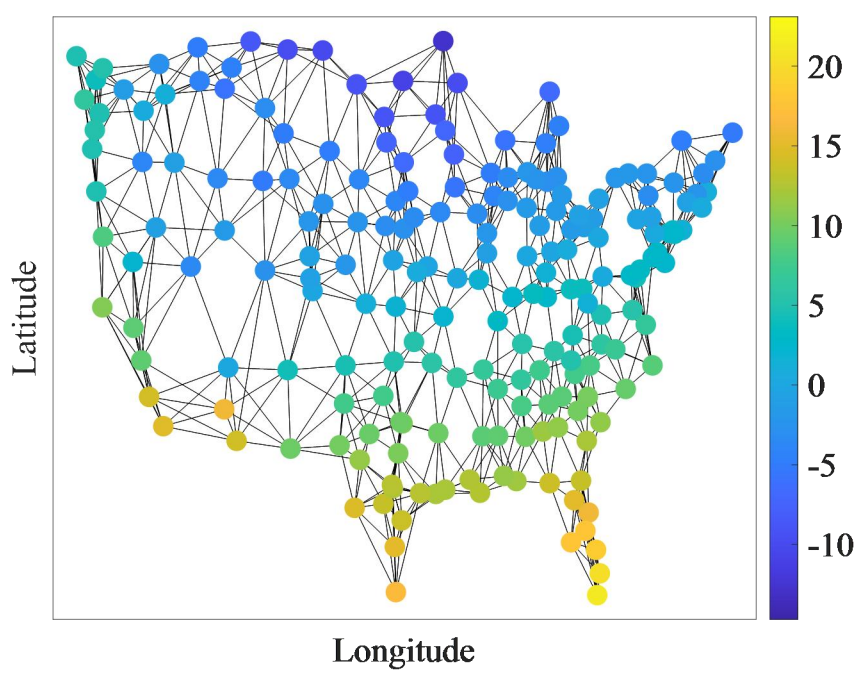

Fig. 13 Graph based on the 12th hour data in the U.S. dataset

Table 9 The related algorithms and their corresponding parameters

| Algorithm | The corresponding parameters |
|---|---|
| GSP LMS | $\mu = 0.5$ |
| GSP NLMS | $\mu = 0.1$ |
| GSP GMCC | $\mu = 0.5,\ \lambda = 0.01,\ \alpha = 1.8$ |
| GSP LOG | $\mu = 0.5,\ \alpha = 0.01$ |
| GSP HQC | $\mu = 0.5,\ \tau = 0.01$ |

The temperature prediction results of the related algorithms in impulsive noise environment are shown in Fig. 13.

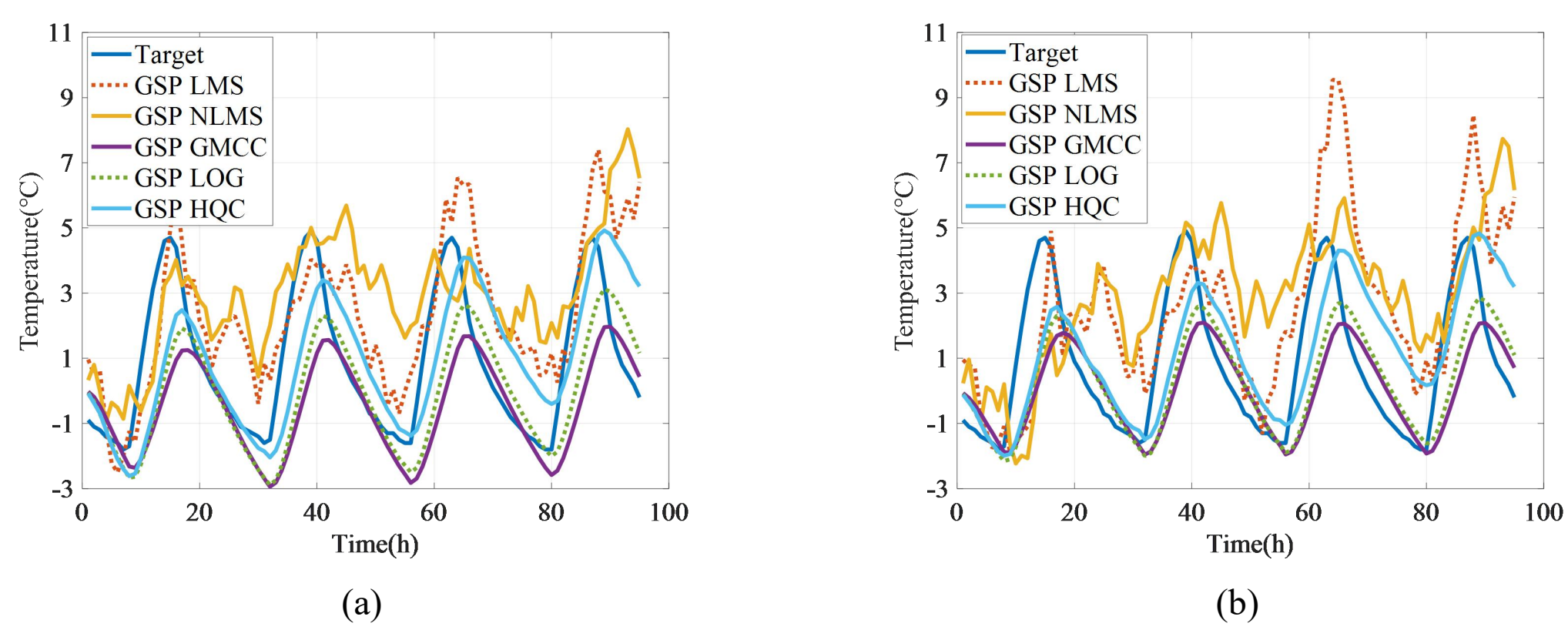

Fig. 13 Temperature prediction results of the related algorithms in impulsive noise environment

(a) $Pr = 0.10$ (b) $Pr = 0.15$

From Fig. 13, it can be found that the temperature prediction curves of the GSP LMS algorithm and the GSP NLMS algorithm fluctuate more violently in the impulse noise environment, and the degree of violence increases with the increase of $Pr$. Meanwhile, the GSP HQC algorithm is closer to the real temperature value compared to the GSP LOG algorithm and the GSP GMCC algorithm.

Fig. 14 shows the algorithms' temperature predictions under Cauchy noise.

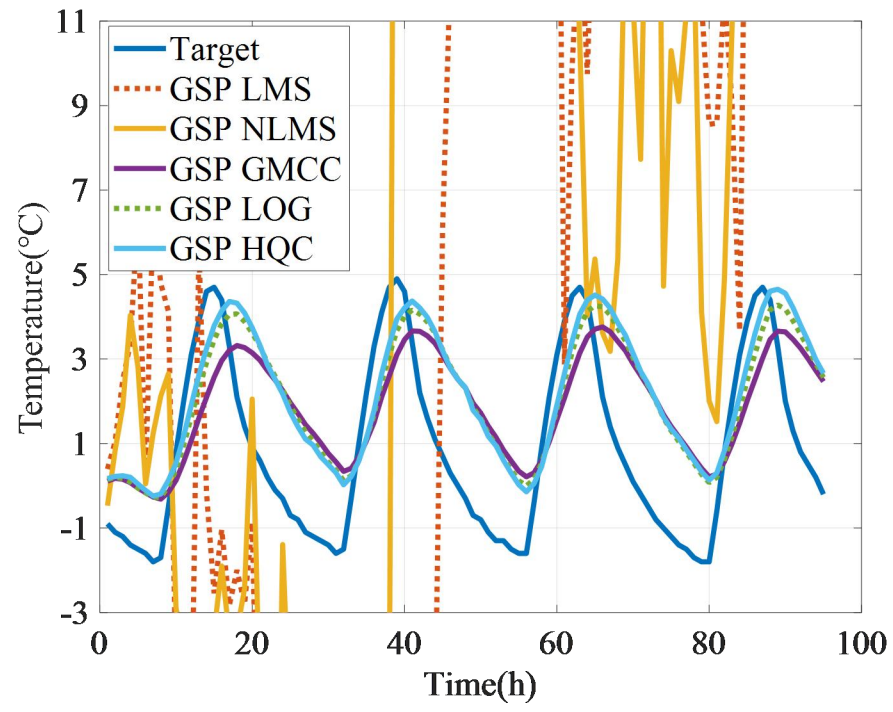

Fig. 14 Temperature prediction results of the related algorithms in Cauchy noise environment

The temperature prediction result of the related algorithms in Laplace noise environment is shown in Fig. 15.

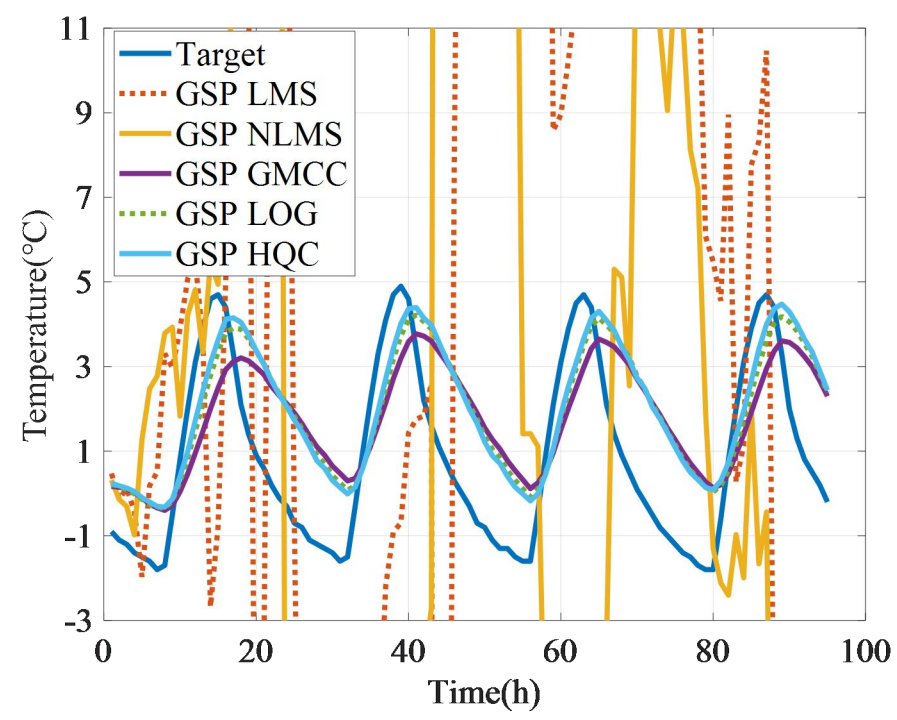

Fig. 15 Temperature prediction results of the related algorithms in Laplace noise environment

From Fig. 14 and Fig. 15, it can be found that the fluctuation program of the temperature prediction curves of the GSP LMS algorithm and the GSP NLMS algorithm in the Cauchy and Laplace noise environments increases dramatically compared to the impulse noise environment, indicating that the prediction effectiveness of the GSP LMS algorithm and the GSP NLMS algorithm in Cauchy and Laplace noise environments is severely deteriorated. Meanwhile, the GSP HQC algorithm is closest to the real target temperature among all algorithms.

Based on the analysis of algorithm computational complexity presented in Section IV, we provide quantitative metrics for the computational complexity of each algorithm within the temperature prediction application, using the complexity of the GSP-LOG algorithm as the benchmark. The results are shown in Table 10.

Table 10 Analysis of Computational Complexity in Temperature Forecasting Applications

| Algorithm | Overall computational complexity of the algorithm | Calculation Results |
|---|---|---|
| GSP LMS | $2F\lvert\mathbf{S}\rvert+2FN-F-N$ | 0.7808% |
| GSP NLMS | $2F^2\lvert\mathbf{S}\rvert+2F\lvert\mathbf{S}\rvert+2F^2N+NF-F^2-F-N+\frac{1}{3}F^3$ | 105.8104% |
| GSP MCC | $2N^2F+N^2+\lvert\mathbf{S}\rvert+2N$ | 99.9981% |
| GSP GMCC | $2N^2F+N^2+4N+\lvert\mathbf{S}\rvert$ | 100.0019% |
| GSP LOG | $2N^2F+N^2+3N+\lvert\mathbf{S}\rvert$ | 100% |
| GSP HQC | $2N^2F+N^2+4N+\lvert\mathbf{S}\rvert$ | 100.0019% |

## 6. Conclusion

In this paper, the advantages of the HQC cost function with strong convexity property are briefly analyzed, and the HQC cost function applicable to graph signal is established, combined with the stochastic gradient descent method, the GSP HQC algorithm is proposed. Secondly, our proposed GSP HQC algorithm is analyzed for its theoretical performance such as steady state error. Again, the better parameters of the GSP HQC algorithm are determined and the correctness of the step size constraint range of the GSP HQC algorithm is verified. Finally, the superior performance of the GSP HQC algorithm in various non-Gaussian noise environments is demonstrated.

The GSP-HQC algorithm proposed in this paper demonstrates favorable performance in experiments; however, it still has two main limitations that require further investigation. First, to obtain closed-form analytical results, several common approximations were adopted in the theoretical derivations, whose accuracy may be limited under strongly non-stationary or highly correlated inputs. Second, the derivations based on second-order moments may become invalid in the presence of heavy-tailed noise. Building upon these limitations, future research could focus on developing theories and practical robust schemes suitable for heavy-tailed noise, designing online adaptive parameter selection methods, and exploring distributed or online implementations to accommodate large-scale, time-varying graphs.

## Data availability

The data that support the findings of this study are available from the corresponding author on request.

## Acknowledgement

This work was partially supported by National Natural Science Foundation of China (grant: 62171388, 61871461, 61571374).

## Statements and Declarations

**Competing Interests**: The authors declare that they have no known competing financial interests or personal relationships that could have appeared to influence the work reported in this paper.

## Appendix :

Table A Symbol/Parameter Quick Reference

| Symbol/ Parameter | Meaning / Definition | Dimension |
|---|---|---|
| $N$ | Total number of nodes in the graph (number of data points) | Scalar |
| $K$ | Number of neighbors per node (for constructing the adjacency) | Scalar |
| $F$ | Bandwidth of the spectral subspace (number of spectral components) | Scalar |
| $\|\mathbf{S}\|$ | Number of observable nodes (sampling points) | Scalar |
| $\mathbf{A}$ | Adjacency matrix | $N \times N$ |
| $\mathbf{\Pi}$ | Diagonal matrix of eigenvalues of the adjacency matrix | $N \times N$ |
| $\mathbf{U}_F$ | Matrix composed of the $F$ eigenvectors corresponding to the frequency set $F$ from the graph Fourier basis, arranged column-wise | $N \times F$ |
| $\mathbf{s}$ | Graph signal represented in the spectral domain | $F \times 1$ |
| $\mathbf{x}$ | Graph signal vector in the vertex domain | $N \times 1$ |
| $\mathbf{D}_\mathbf{s}$ | Sampling matrix (diagonal matrix containing only 0 or 1) | $N \times N$ |
| $\mathbf{B}$ | Bernoulli diagonal matrix (indicator for impulse noise occurrence) | $N \times N$ |
| $Pr$ | Probability of impulse noise occurrence (Bernoulli parameter) | Scalar |
| $\delta_\eta^2, \delta_\gamma^2$ | Noise variance | Scalar |
| $\tau$ | Design parameter of the HQC cost function | Scalar |
| $\mu$ | Step size | Scalar |
| $\lambda$ | Kernel width in MCC/GMCC | Scalar |
| $\alpha$ | Shape parameter in GMCC/LOG | Scalar |
| $\mathbf{e}$ | Error vector | $N \times 1$ |
| $\mathbf{G}(\mathbf{e})$ | Error handling function (diagonal matrix) | $F \times F$ |
| $\mathbf{K}$ | Fixed non-negative weight matrix (size F×F) introduced for steady-state error analysis | $F \times F$ |
| $\mathbf{T} = vec(\mathbf{T})$ | Intermediate matrices and their vectorized forms used in derivations | $F \times F$ |
| $\mathbf{H}, \mathbf{R}$ | Intermediate matrix arising in mean-square derivation | $F \times F$ |
| $k$ | Multiple of the critical value used to set the step size in experiments | Scalar |